\documentclass[final]{siamltex}

\usepackage{amsmath,amsfonts,amssymb,array}
\usepackage{color,wrapfig}
\usepackage{graphicx}

\def \beq {\begin{equation}}
\def \eeq {\end{equation}}
\def \ba {\begin{eqnarray}}
\def \ea {\end{eqnarray}}
\def \ban {\begin{eqnarray*}}
\def \ean {\end{eqnarray*}}
\def \bfl {\begin{flalign*}}
\def \efl {\end{flalign}}
\def \bsp {\begin{split}}

\def \l {\left}
\def \r {\right}

\newcommand{\upp}{\hspace{-0.2 pt}\uparrow}
\newcommand{\downn}{\hspace{-0.2 pt}\downarrow}

\newcommand{\ketbrad}[1]{|#1\rangle\!\langle #1|}
\newcommand{\ketbradt}[2]{|#1\rangle\!\langle #2|}
\newcommand{\ketbrads}[1]{|\!#1\rangle\!\langle #1\!|}
\newcommand{\ketbradts}[2]{|\!#1\rangle\!\langle #2\!|}

\newcommand{\mb}[1]{\mbox{\boldmath$#1$}}
\newcommand{\mbb}[1]{\mathbb{#1}}

\newcommand{\mean}[1]{\langle#1\rangle}

\def\ket#1{\left| #1\right>}
\def\bra#1{\left< #1\right|}
\newcommand{\poly}{\textnormal{poly}}

\title{Algorithmic Cooling of a Quantum Simulator}
\author{Dvir Kafri\thanks{Joint Quantum Institute, University of Maryland, College Park, MD} \and Jacob M. Taylor\thanks{Joint Quantum Institute, University of Maryland, College Park, MD \and National Institute of Standards and Technology, Gaithersburg, MD}}

\begin{document}

\maketitle

\begin{abstract}
  Controlled quantum mechanical devices provide a means of simulating more complex quantum systems exponentially faster than classical computers.  Such ``quantum simulators'' rely heavily upon being able to prepare the ground state of Hamiltonians, whose properties can be used to calculate correlation functions or even the solution to certain classical computations.  While adiabatic preparation remains the primary means of producing such ground states, here we provide a different avenue of preparation: cooling to the ground state via simulated dissipation.  This is in direct analogy to contemporary efforts to realize generalized forms of simulated annealing in quantum systems.
\end{abstract}



\pagestyle{myheadings}
\thispagestyle{plain}
\markboth{ Algorithmic Cooling of a Quantum Simulator}{ D. Kafri and J. M. Taylor}

\section{Introduction}
Quantum devices provide new opportunities in communication and computation~\cite{Feynman1982,Lloyd1996,BB84,E91,Bennet1992,Grover1997,Shor1999,Berzina2004,Harrow2009}.  One promising application of a well controlled quantum device is simulating a quantum system, which can occur exponentially faster than can be achieved classically~\cite{Lloyd1996,Terhal2000,Biamonte2011,Weimer2010}. Such simulations could provide insights in many current fields of research, such as BCS-BEC superfluids \cite{Bakr2010,Bloch2012}, quantum chemistry \cite{Kassal2008,Kassal2011}, and highly correlated condensed matter systems \cite{Auerbach1994,Fabian2005,Lee2006,Lewenstein2007}. However, a crucial component of such simulation is the specification of the initial state of the system to be simulated.  While for some problems, including many in quantum chemistry \cite{Aspuru-Guzik2005}, such initial states can be found via prior knowledge from classical computer studies, in general a means of preparing such states does not exist. 

Methods for the preparation of specific eigenstates of Hamiltonians, particularly the ground state, therefore remain a pressing challenge for the most interesting quantum simulation applications.  One approach for preparation is adiabatic evolution from a system with an accessible ground state~\cite{Farhi2000}.  Here, we offer an alternative approach: cooling to the ground state by expanding the simulation to include a (small) quantum bath.  This approach differs crucially from prior work in several respects.  In contrast with adiabatic approaches \cite{Dam2001}, it requires only information concerning the spectrum of the cooled Hamiltonian $H_S$ and not any intermediate Hamiltonians of the form $\lambda H_S + (1 - \lambda) H_0$. In its most general form it is able to prepare ground states of a wide class of gapped Hamiltonians. Specifically, it may cool any gapped Hamiltonian with a tight band of excited state energies that are separated from the rest of the spectrum.  For appropriate cases, we find our approach provides a quadratic speedup for ground state preparation in analogy with Grover's algorithm \cite{Grover1996}. Finally, the approach may also be used to cool a variant of Kitaev's clock Hamiltonian \cite{Kitaev2003} at a polynomial overhead, so that it may efficiently solve any problem solveable through conventional circuit-based quantum computation \cite{Nielsen2000}.  

\section{Grover's Algorithm by Simulated Cooling}
As an illustrative example, we apply the method of  QSC (Quantum Simulated Cooling) to the Hamiltonian analogue of Grover's Algorithm \cite{Grover1996}. Although the arguments given in this section are heuristic, the claims made for the general scheme are rigorously shown in the supplemental. Say that we are given a function $f: \{0,1 \}^n \rightarrow \{0,1 \}$, and we wish to find $y\in \{0,1 \}^n $ such that $f(y) = 0$. Analogously, say that we are able to simulate a Hamiltonian $H_S$ on $n$ qubits, which has only two distinct eigenspaces. These are labeled $P_{0}$ and $P_1$, with energies $\omega_0 = 0$ and $\omega_1$, and correspond to the logical qubit states $\ket{x}$ such that $f(x) = 0$ (1), respectively. For notational simplicity, we let the symbol for an eigenspace also represent its projector. We then have
\ba
\label{HS1}
H_S & = & \omega_1 P_1
\ea
Up to a constant factor, the simulation of $H_S$ for a fixed time is equivalent in cost to evaluation of $f(x)$ \cite{Nielsen2000}, so solving for $y$ is equivalent to finding a state in the zero energy manifold of \eqref{HS1} \cite{Farhi1998,Nielsen2000}.

To take the system from an initial state $\ket{F}$ to the ground state manifold, we concurrently evolve a single qubit bath, with Hamiltonian $H_B = \omega_B \mb{1}_S\otimes \ketbrads{\upp}$, and define the non-interacting Hamiltonian $H = H_S + H_B$. We prepare the system and bath in the state $\ket{F \downn}$, and introduce a coupling between them, denoted by 
\ba
\label{V}
V = \Omega_0 \ketbrad{F} \otimes \sigma_x
\ea
To prevent accidental symmetries leading to frustration, $\ket{F}$ is a randomly generated quantum state \cite{Gross2007}. 

 To illustrate the evolution of our state with this interaction, we decompose $\ket{F}$ into its spectral components:
\ba
\ket{F} & = & x_0 \ket{0} + x_1 \ket{1}
\ea
where $x_j \ket{j} = P_j \ket{F}$ and $x_j$ is real and non-negative. As the space $\mb{S} = $ \\$\mbox{Span}\{\ket{0 \downn}, \ket{0, \upp}, \ket{1,\downn},\ket{1, \upp}\}$ is invariant under $H+V$, if we prepare the system in the state $\ket{F}$, we may reduce our analysis to this subspace. Written explicitly, within $\mb{S}$ the full Hamiltonian takes the form
\ban
 \left(\begin{array}{l |c c c c}
 & (0 \downn) & (0 \upp) & (1 \downn) & (1 \upp)\\
\hline
(0 \downn) & 0 & \Omega_0  x_0^2         & 0         & \Omega_0 x_0 x_1       \\
(0 \upp)   & \Omega_0  x_0^2  & \omega_B  & \Omega_0 x_0 x_1            & 0              \\
(1 \downn) & 0 & \Omega_0  x_0 x_1           & \omega_1    & \Omega_0 x_1^2       \\
(1 \upp)   & \Omega_0 x_0 x_1    & 0         & \Omega_0 x_1^2  & \omega_1+\omega_B
\end{array}\right)
\ean

Suppose that we set $\Omega_0 \ll \omega_1$, and $\omega_B \approx \omega_1$. Since $\ket{0 \upp}$ and $\ket{1 \downn}$ are nearly degenerate, we expect that superpositions of these states form eigenstates of $H+V$. As $V$ has no diagonal terms in the above basis, we see that even orders of $\Omega_0$ induce level shifts on these states, while odd orders couple between them. With knowledge of $x_0$ and $x_1$, we could compute \cite{Cohen-Tannoudji1992} the even-order level shifts induced by $V$ on the energies of $\ket{0\, \upp}$ and $\ket{1 \, \downn}$, and adjust $\omega_B$ accordingly so that they are degenerate. To leading order, determining $\omega_B$ is equivalent to solving $\omega_B +\frac{ x_0^4 \Omega_0^2}{\omega_B} = \omega_1 + \frac{x_1^4\Omega_0^2}{\omega_B}$. Notice that for $|x_0|\ll |x_1|\approx 1$, the level shift of state $\ket{1 \downn}$ is approximately $\frac{ x_1^4 \Omega_0^2}{\omega_B}$, which is non-negligible compared to the coupling $\Omega_0 x_0 x_1$. As seen in Figure 1, if this shift is not accounted for the scheme's success rate becomes exponentially small with increasing $n$. 
\begin{figure}
\includegraphics[scale= 0.43]{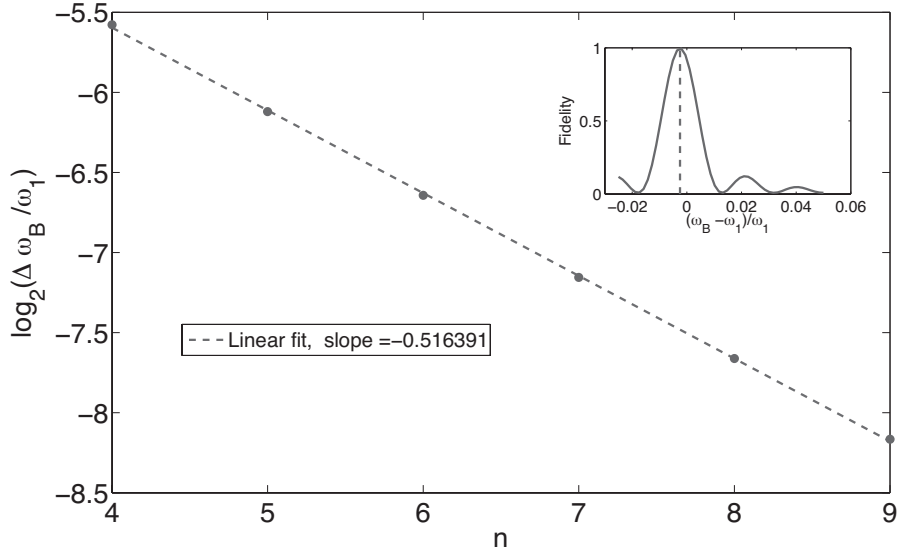}
\caption{This figure illustrates the sensitivity of the Grover cooling scheme to errors in bath detuning, as a function of $n$. As seen in the inset for $n = 7$, the probability of the cooling scheme to produce the ground state can be computed as a function of bath detuning, $(\omega_B-\omega_1)/\omega_1$. The vertical dashed line represents the first order correction to the level shift between states $\ket{1 \downn}$ and $\ket{0 \upp}$, which is $\Omega_0^2/\omega_1$ for $|x_1|\approx 1$. The main figure plots the half-width of this fidelity function, for $\Omega_0/\omega_1 = 0.05$.  The dashed line represents a linear fit, suggesting that the half-width scales approximately as $\sqrt{2^{-n}}\omega_1/\Omega_0$. Naively setting $\omega_B = \omega_1$ requires that, for a fixed success probability, $\Omega_0/\omega_1$ must be scaled as $\sqrt{2^{-n}}$, thereby negating the quadratic speedup observed with the optimal detuning. }
\end{figure}

Since the interaction strength $\Omega_0$ is perturbative, in the limit $|x_0|\ll 1$ we conclude that for the manifold of states with energy near $\omega_B$, the Hamiltonian is effectively
\ban
H_{eff} &=& \omega_B \left( \ketbradt{0 \upp}{0 \upp\!} + \ketbradt{1 \downn}{1  \downn\!} \right) \\
& \quad & + \Omega_0 x_0 x_1 \left( \ketbradt{0 \upp}{1 \downn\!} + \mbox{h.c.} \right)
\ean
while the states $\ket{0 \, \downn}$ and $\ket{1\, \upp}$ are unchanged by $H+V$.  We thus observe coherent oscillations between $\ket{0 \, \upp}$ and $\ket{1 \, \downn}$, at a rate $\Omega = \Omega_0 x_0 x_j$. If we prepared the system and bath in the state $\ket{F \downn}$, we could evolve for a time $\tau = \frac{\pi}{2 \Omega}$. The resulting output would be
\ban
e^{- i \tau (H + V) } \left(x_0 \ket{0\, \downn} + x_1 \ket{1\, \downn} \right) \approx x_0 \ket{0\, \downn} + x_1 \ket{0 \upp}
\ean
so that the system is in a groundstate of $H_S$. 

We can relate the actual cost of the algorithm to the simulation time by noting that a single implementation of $\exp(-i t H_S)$ is equivalent in cost to $O(1)$ evaluations of the function $f$ \cite{Nielsen2000}. Using a stroboscopic expansion \cite{Suzuki1990}, simulation of  $H_S + H_B + V$ for total time $T$ may then be done on a standard quantum computer at a cost approaching $O(||H||T)$ \cite{Berry2007}. Since the time scale necessary to map between $\ket{1 \, \downn}$ and $\ket{0 \, \upp}$ is set by the Rabi rate $\Omega = \Omega_0 x_0 x_1$, one sees that for fixed $\omega_1$, the total cost of the algorithm scales linearly with $|x_0|^{-1}$.  The scaling of simulation time as $|x_0|^{-1}$ will also apply to generalizations of QSC to more complicated Hamiltonians. 

Say that $N = 2^n$ is the dimension of the $n-$qubit Hilbert space, and $N_0 = \mbox{dim } P_0$. If we select $\ket{F}$ from a random sample, so that on average $|x_0|^2 \approx N_0/N$, the running time of the algorithm will scale as $\sqrt{N/N_0}$, reflecting the quadratic speedup over classical computation observed in Grover's Algorithm. On a standard quantum computer, such a sampling can be achieved through $\epsilon$-approximate unitary 2-designs, which may be implemented at a cost of $O(n \log(1/\epsilon))$ \cite{Dankert2006}. Note that being able to set $\tau$ correctly, as well as correcting for the level shift induced by $V$, requires knowledge of the value of $x_0$ and $x_1$. This issue is also relevant to the more general problem, and in the case where the decomposition of $\ket{F}$ is unknown, we present a modified scheme below that succeeds probabilistically in the same time.

\begin{figure*}
\includegraphics{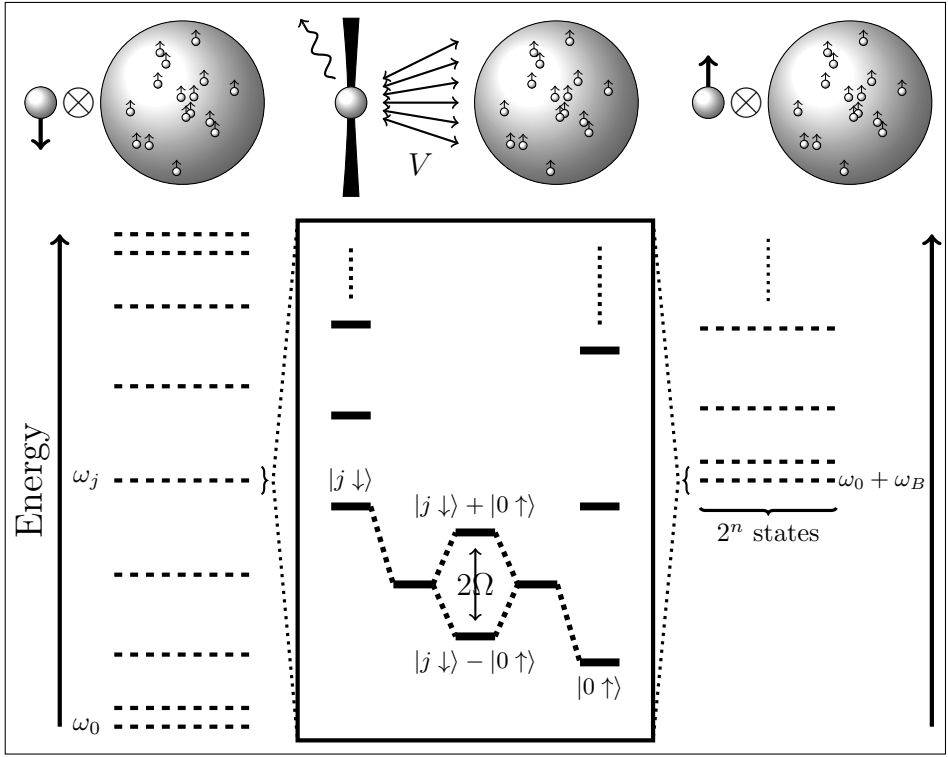}
\caption{The spectrum of $H_S + H_B$ during the cooling of energy level $j$. The left and right columns represent the spectrum of the composite system when the bath spin is in its ground and excited state, respectively. Each band (dashed line) represents a degenerate subspace of $H_{prop}$, whose degeneracy is lifted by $H_{input}$. The bath gap $\omega_B$ is adjusted so that the ground state band on the right is nearly degenerate with band $j$ on the left. The inset (center column) shows the effect of the interaction $V$ on the two unperturbed eigenstates in $\mb{S}_1$, $\ket{j \downn}$ and $\ket{0 \upp}$. The level shift in these states is due to even order corrections in $V$, and the value of $\omega_B$ is adjusted so that these shifted states are exactly degenerate. The odd order corrections in $V$ then allow for coherent oscillations between these states at frequency $2 \Omega$.  After interacting for time $\pi/(2 \Omega)$, the bath spin is pumped back to its ground state and the process is repeated for band $j-1$.}
\end{figure*}

\section{Cooling a Quantum Circuit}
Here we show how QSC can be used to produce the outcome of a chain of 2-qubit unitary operations, $\mb{U} = U_L U_{L-1} ... U_{1}$, at a cost scaling as $O(\poly(L))$. Since 1 and 2-qubit unitaries are sufficient to implement any efficient quantum computation \cite{Barenco1995,Lloyd1995}, any problem efficiently solved through standard quantum computation can also be solved using QSC with at most a polynomial overhead. The idea behind our result draws from the work of \cite{Aharonov2007}, which shows that adiabatic quantum computation is equivalent to standard quantum computation. 

Suppose there exists a Hamiltonian $H_S$ whose unique groundstate, after tracing out any ancilla qubits, can be made arbitrarily close to $\mb{U}\ket{0^n}$. Preparing the groundstate of $H_S$ would correspond to producing the outcome of the computation. One $H_S$ satisfying this requirement is a variant of Kitaev's clock Hamiltonian \cite{Kitaev2002}. As in \cite{Feynman1986}, to describe $H_S$ we consider a particle living on a 1D lattice with $L + 1$ sites, whose internal state is described by $n$ qubits. For a given site $l$, the particle has fixed onsite energy $\omega$, but may also tunnel to neighboring site $l+1$ through a coupling term $-\frac{\omega}{2}\cdot U_l$ acting on the internal states. The Hamiltonian describing the particle is then
\ba
H_{prop}/w = \sum_{l = 0}^L \ketbrad{s_l} - \frac{1}{2}\sum_{l=1}^L \left( U_l \ketbradt{s_l}{s_{l-1}} + \mbox{h.c.}\right)
\ea
where $\ket{s_l}$ corresponds to the particle being in the $l$th site. This Hamiltonian is analogous to that of a particle freely propagating through space, and its eigenstates are all of the form \cite{Kempe2004}:
\ba
\label{eigens}
c_0\ket{x}\ket{s_0} + \sum_{l=1}^L c_l \l( U_{l}U_{l-1}...U_{1}\r)\ket{x}\ket{s_l} 
\ea
where $\ket{x}$ is the internal state of the particle at site $l= 0$.

Since we want $\ket{0^n}$ at the start of the computation, we add a perturbation to $H_{prop}$ of the form
\ba
H_{input} = \Delta_1 \sum_{m=1}^n \ketbrad{1}_m \otimes \ketbrad{s_0}
\ea 
where $\Delta_1\ll \omega$, and $\ketbrad{1}_m$ acts only on qubit $m$ of the particle's internal state. This lifts the degeneracy between eigenstates with different initial conditions, and allows us to reduce our analysis to the invariant subspace of states with $\ket{x} = \ket{0^n}$ in \eqref{eigens}. We label this subspace $\mb{S}_1$ and its complement $\mb{S}_2$.

The ground state of $H_{prop}$ in $\mb{S}_1$ corresponds to $c_l = \frac{1}{\sqrt{L+1}}$ for all $l$. The site $L$ component of the ground state is $\frac{1}{\sqrt{L}}\mb{U}\ket{0^n}\ket{s_L}$, so by preparing this state and measuring the particle in $L$ we would obtain the outcome of the computation. If we instead concatenate $O(L/\epsilon)$ identity operations to the definition of $H_{prop}$, then after tracing out the particle's position its internal state has an $O(\epsilon)$ trace-norm distance from $\mb{U}\ketbrad{0^n}\mb{U}^\dagger$ \cite{Aharonov2007}.  

The eigenstates of $H_S = H_{prop} + H_{input}$ in $\mb{S}_1$ are non-degenerate and have energy $\omega_j = \omega\cdot \left(1  -\cos\left( \frac{j \pi}{L+1}\right) \right)$, where $0\leq j \leq L$. There are two other important energy scales associated with $H_S$. The first is the maximal energy, $||H_S|| = \omega\cdot O(L)$ \cite{Aharonov2007}, where $||\cdot ||$ is the spectral norm. The other relevant energy is the spectral resolution of $\mb{S}_1$: 
\ba
\Delta = \min_{\omega_j,E}\{ |E - \omega_j| : E \in \mbox{Spec}(H_S), E\neq \omega_j \}
\label{Delta}
\ea
As seen below, the ratio $||H_S||/\Delta$, where $|| \cdot ||$ is the operator norm, will determine the overall cost of the simulation. Further, for a given error tolerance the energy scale $\Delta$ provides an upper bound to the system-bath coupling $V$ and allowable error terms in the sumulation. 

To bound $\Delta$, we note that the eigenspaces of $H_{prop}$ are degenerate bands corresponding to the spectrum $\{\omega_j\}_{j=0}^L$, which are indexed by the state $\ket{x}$ in \eqref{eigens}.  $H_{input}$ is diagonal within each $\omega_j$ band, with diagonal entries $\Delta_1 N(x) h_j$, where $N(x)$ is the number of $1$'s in the binary expansion of $\ket{x}$. By diagonalizing $H_{prop}$ \cite{Yueh2005}, one may verify that $1/h_j=O(L^3)$. Since $H_{input}$ vanishes only when $\ket{x}=\ket{0^n}$, as long as first order degenerate perturbation theory is valid we see that the spectra of $H_S$ in $\mb{S}_1$ and $\mb{S}_2$ are distinct. This is true when $||H_{input}|| = n \Delta_1$ is bounded by the minimum level spacing of  $H_{prop}$, which scales as $\omega\cdot  L^{-2}$.  We therefore assume that $\Delta_1 \propto \omega\cdot n^{-1} L^{-2} $, so the gap satisfies $\Delta^{-1} \leq (\Delta_1 h_j)^{-1} = \omega^{-1} \cdot O(n L^5)$.  

To prepare the groundstate of $H_S$, we simulate a single qubit bath coupled to the system, with energy splitting $\omega_B$. The projectors into $\mb{S}_1$ and $\mb{S}_2$ act trivially on this qubit.  The fiducial state is $\ket{F \downn} = \ket{0^n}\ket{s_0}\ket{\downn} \in \mb{S}_1$. We also introduce the system-bath interaction $V = \Omega_0 \mb{1}_n\otimes\ketbrad{s_0} \otimes \sigma_x$. Like $H_S$, $V$ is block diagonal in $\mb{S}_1$ and $\mb{S}_2$ and satisfies
\ba
\label{V2}
\mb{S}_1 V \mb{S}_1= \Omega_0 \ketbrad{F}\otimes \sigma_x
\ea 
Letting $\ket{j}$ denote the $j$-th excited state in $\mb{S}_1$, we write the spectral decomposition of $\ket{F}$:
\ba
\label{Fdecomp}
\ket{F} = \sum_{j} x_j \ket{j}
\ea
Although the states $\ket{j}$ are dependent on $\mb{U}$, the coefficients $x_j$ are only dependent on $L$ and may be calculated explicitly \cite{Yueh2005}.

The algorithm proceeds as follows. We start with the bath energy $\omega_B$ near $\omega_L$, to address the transition $\ket{L \downn} \rightarrow \ket{0 \upp}$ (see Figure 2). In order for this to be favorable, we must calculate the even-order level shifts induced by $V$ on states $\ket{0 \upp}$ and $\ket{L \downn}$, and adjust $\omega_B$ so that they are degenerate. Such a calculation is equivalent to finding a root of a degree $L$ polynomial to accuracy $1/\poly(L)$, and is possible since one may explicitly calculate the coefficients $x_j$ in \eqref{Fdecomp}. We evolve the system and bath under the Hamiltonian $H+V$ for a time $\tau_L = \frac{\pi}{2 \Omega}$, where $\Omega = \Omega_0 |x_0 x_L|(1 + O(\Omega_0/\Delta))$ is the coupling rate between $\ket{0\upp}$ and $\ket{L \downn}$. This maps the component of $\ket{F \downn}$ in the state $\ket{ L \downn}$ to the state $\ket{0 \upp}$, while leaving the lower energy space $\mbox{Span} \left\{\ket{j \downn}\right\}_{j=0}^{L-1}$ unchanged. We then measure the bath in the logical basis. A measurement of $\ket{\upp}$ implies that the desired transition has occurred, so we may terminate the algorithm. A measurement of $\ket{\downn}$ implies that we have projected to $\mbox{Span} \left\{\ket{j \downn}\right\}_{j=0}^{L-1}$. We therefore decrement $\omega_B$ to a value near $\omega_{L-1}$, account for the level shifts of $\ket{0 \upp}$ and $\ket{(L-1) \downn}$, evolve for time $\tau_{L-1}$, and remeasure the bath. Repeating this process for at most $L$ evolutions, we reach the ground state with high probability.

To prove this claim, we use the language of trace-preserving, completely positive (TCP) maps \cite{Choi1975, Kraus1971, Bengtsson2008}. The map associated with measurement and conditional evolution at energy near $\omega_j$ is
\ba
\label{TCP1}
\begin{split}
E_j(\rho) &=  U_j \ketbrads{\downn } \rho \ketbrads{\downn} U_j^\dagger  \\
& \quad +  \ketbrads{\upp} \rho  \ketbrads{\upp}
\end{split}
\ea
where $\rho$ is a density matrix for the system and bath, $U_j$ is the unitary evolution under $H+V$ for $\omega_B \approx \omega_j$ and time $\tau_j \approx \frac{\pi}{2 \Omega_0 |x_0 x_j|}$, and $\ketbradt{\!\!\downn}{\downn\!\!} = \mb{1}_S \otimes \ketbradt{\!\!\downn}{\downn\!\!}_B$. We define the TCP map describing the complete algorithm as $\mathbb{E} = E_1 \circ E_2 ... \circ E_L$. In the supplemental section, we build upon results from \cite{Kempe2004,Stewart1990} to show that for $\rho = \ketbradt{F \downn}{F \downn\!}$, 
\ba
\begin{split}
 \mathbb{E}(\rho) &= \lambda_0 \rho_0 + \hat R
\end{split}
\ea
where $\lambda_0\geq 0$ and $\rho_0$ is a density matrix made up of states $\ket{0 \downn}$ and $\ket{0 \upp}$. $\hat R$ represents an error term satisfying $\mbox{Tr}[\hat R] = O(L^{5/2} \Omega_0/\Delta) = ||\hat R||$. For the algorithm to succeed with error rate $\epsilon = \mbox{Tr}[ (\mb{1}- \ketbrad{0}) \mathbb{E}(\rho)] = \mbox{Tr}[\hat R] - \mbox{Tr}[\ketbrad{0} \hat R] $, we must scale $\Omega_0$ as $ O(\epsilon \Delta L^{-5/2} )$. 
\subsection{Timing and Cost}
We now discuss the relationship between the simulation time $T$ and the actual cost of implementing the algorithm. We assume that the cost of simulating $n$ qubits under a Hamiltonian $H$ for time $T$ scales as $\poly\l(||H|| T\r)$. This is true for all $H$ that may be written as a sum of $\poly(n)$ interactions, each involving a fixed number of qubits \cite{Lloyd1996}. For our clock Hamiltonian, we may implement $H_S+ H_B + V$ using at most 5-qubit interactions \cite{Aharonov2007}, and expect that a 2-local QSC scheme should also be possible using methods in \cite{Jordan2008, Bravyi2008}.

To leading order in $(\Omega_0/\Delta)$ the time required for cooling step $j$ is $\tau_j = \frac{\pi}{2 \Omega_0 |x_0 x_j|}$, so the total time of the algorithm may be computed as 
\ba
\label{time1}
T = \sum_{j = 1}^{L} \tau_j = \frac{\pi}{2 \Omega_0 |x_0|} \sum_{j=1}^L \frac{1}{|x_j|}
\ea
By diagonalizing $H_{prop}$ within the space $\mb{S}_1$, a simple calculation shows that the coefficients $x_j$ in \eqref{Fdecomp} satisfy $|x_j|^{-1} = O\left(\sqrt{L}/\cos\left(\frac{j \pi}{2(L+1)} \right) \right)$. Since $\Omega_0 = O(\epsilon \Delta L^{-5/2})$, the total simulation time $T$ scales as $\Delta^{-1} \cdot O\left(L^5 \sqrt{L} \log(L)/\epsilon \right) $. Finally, as $\Delta = O(n^{-1}L^{-5})$ and $||H|| = O(L)$, the actual cost of the algorithm is then set by $||H||T = O\left(n L^{10}\sqrt{L} \log(L)/\epsilon \right)$. QSC is therefore equivalent to standard quantum computation: any problem on $n$ qubits that may be solved in $\poly(n)$ time using a standard quantum computer may also be solved in $\poly(n)$ time through QSC.  

\section{Extension}
Although the scheme above is in principle as powerful as circuit-based quantum computation, its applicability towards cooling other Hamiltonian systems is limited by the assumptions placed on $H_S$ and $V$. To avoid unwanted transitions, it requires an accessible non-degenerate eigenspace $\mb{S}_1$ with gap $\Delta$, as well an interaction of the form \eqref{V} described by a known decomposition \eqref{Fdecomp}. Here we propose an extension to this scheme that overcomes these constraints. The caveats of this approach are that it is probabilistic and requires $(k+1)$-local interactions for $k$-local $H_S$. 

We briefly summarize the scheme before describing it in detail. We assume that the spectrum of $H_S$ has a non-degenerate ground state $\ket{0}$, as well as a manifold $P_1$ of energies near $\omega_1$ that is well separated from the rest of the spectrum. We also assume the ability to simulate evolution under to an operator $T_S$ that couples between this manifold and $\ket{0}$. We define Hamiltonians $H$ and $V$ based on $H_S$ and $T_S$ such that coherent oscillations occur between $\ket{1 C}$ and $\ket{0 B}$, where $\ket{1}\in P_1$ and $\ket{B}$,$\ket{C}$ are states of a qutrit bath. We start with a fiducial state $\ket{F C}$ and evolve under $H+V$ for a sufficiently long time, after which we check if the bath is in state $\ket{B}$. If that is the case, we verify that the system has transitioned to the ground state by mapping $\ket{B}\rightarrow \ket{L}$ and evolving under a verification Hamiltonian \eqref{Hver}. A bath measurement of $\ket{R}$ heralds the success of the scheme, and otherwise we repeat the process.

We now specify the scheme's requisite assumptions. First, the Hamiltonian $H_S$ is of the form
\ba
\label{HSextension}
H_S = P_1 H_S P_1 + P_2 H_S P_2
\ea
where $P_1$ and $P_2$ are projectors into eigenspaces of the same name. The eigenspace $P_1$ represents a narrow band of energies within $(\omega_1-\delta \omega_1,  \omega_1+\delta \omega_1)$. We assume that $H_S$ has a non-degenerate ground state $\ket{0}$ with energy $\omega_0 = 0$, and let $P_2$ represent the space orthogonal to both $P_1$ and $\ket{0}$. The limiting energy scale for this scheme is $\Delta = \min\l\{\omega_1, E, |E-\omega_1| : E \in\mbox{Spec}(H_S|_{P_2})  \r\}$. $\Delta^{-1}$ will set an upper bound for the time scale of the simulation. 

Along with $H_S$, we assume access to a Hermitian operator $T_S$ that couples $\ket{0}$ to the space $P_1$:
\ba
\label{coupling}
P_1 T_S \ket{0} = \Omega \ket{1}
\ea
where $\Omega$ is real and positive by choice of phase convention. Using this coupling, over a time scale $1/\Omega$ the simulation will cause coherent oscillations of the form $\ket{1 C}\leftrightarrow \ket{0 B}$, where $\ket{C}$ and $\ket{B}$ are orthogonal bath states. We must further assume that the energy spread of $P_1$ is small compared to the coupling: $\delta \omega_1 \ll \Omega$. This will allow us to treat $\ket{1}$ as an eigenstate of $H_S$ in our analysis of the simulated evolution. We must also have that the energy shift induced by $T_S$ on the ground state is small compared to the coupling: $|\bra{0}T_S\ket{0}|^2/\omega_1 \ll \Omega$, that $T_S$ is perturbative compared to the gap: $\Omega<||T_S||\ll \Delta$, and that $T_S$ does not couple strongly between $\ket{0},P_1$ and $P_2$: $||(\ketbrad{0} + P_1) T_S P_2|| \ll \sqrt{\Omega \Delta}$.

The original scheme succeeds by introducing a 2-level bath and adjusting its energy $\omega_B$ so that $\ket{1 \downn}$ and $\ket{0 \upp}$ are nearly degenerate. Within degenerate perturbation theory, the interaction $V$ then produces a splitting of $2 \Omega$ between approximate eigenstates $\frac{1}{\sqrt{2}}( \ket{1 \downn} \pm \ket{0 \upp})$, thereby causing coherent oscillations of the form $\ket{1 \downn} \leftrightarrow \ket{0 \upp}$. If not accounted for, the (even-order) level shifts induced by $V$ on the state $\ket{1 \downn}$ can be much larger than this splitting, meaning the eigenstates look more like $\ket{1 \downn}$ and $\ket{0 \upp}$, so that the desired oscillation does not occur. Although we may explicitly account for the level shifts by adjusting $\omega_B$, this requires knowledge of the coefficients $x_j$ in \eqref{Fdecomp}. Instead, one may tailor the unperturbed Hamiltonian $H$ so that unwanted level shifts cancel out:
\ba
\begin{split}
H & = H_S \otimes \left(\ketbrad{C} + \ketbrad{R} - \ketbrad{L}\right) \\
& + \omega_1 \mb{1}_S \otimes \left( \ketbrad{R} + \ketbrad{L} \right)
\end{split}
\ea
The bath Hilbert space $\mathbb{H}_B$ now has dimension 3, with basis vectors $\ket{C}, \ket{R}$ and $\ket{L}$. 

We use the operator $T_S$ to create the system-bath interaction $V$:
\ba
\label{V4}
\begin{split}
V  &=  T_S \otimes \left(\ketbradt{C}{B} + \ketbradt{B}{C}\right) \\
 \ket{B} & =  \frac{1}{\sqrt{2}}\left(\ket{L} + \ket{R} \right)
\end{split}
\ea
To see how $H+V$ causes the desired oscillations, we compute the level shift operator $\Sigma(z)$ for the manifold $P$ of eigenstates of $H$ with energy at most $\Delta/4$ away from $\omega_1$ \cite{Cohen-Tannoudji1992}. Written as a projector, 
\ba
P = \ketbrad{0}\otimes (\ketbrad{L} + \ketbrad{R}) + P_1 \otimes \ketbrad{C}
\ea
$\Sigma(z)$ characterizes the time evolution of the states in $P$ under Hamiltonian $H+V$. As seen in \cite{Aharonov2007} and further developed in the supplemental, the eigenstates and eigenvectors of $\Sigma(z)$ for $z$ close to $\omega_1$ are good approximations of eigenstates and eigenvectors of $H+V$. Written explicitly, we have 
\ba
\begin{split}
\label{Sigma}
\Sigma(z) &= P H P + P V P + P V G(z) V P\\
& \quad + P V G(z) V G(z) V P + ...
\end{split}
\ea
where $G(z) = \frac{Q}{z Q - Q H Q}$ and $Q$ is the complement of $P$. We observe that for $z =\omega_1$, 
\ba
\begin{split}
G(\omega_1) &= \frac{P_1+P_2}{P_1 H_S P_1 + P_2 H_S P_2}\otimes (\ketbrad{L}-\ketbrad{R})\\
&\quad + \omega_1^{-1}\ketbrad{0\, C} + \frac{P_2}{ P_2 H_S P_2} \otimes \ketbrad{C}
\end{split}
\ea
so that $\bra{B}G(\omega_1)\ket{B} = 0$. Thus \eqref{Sigma} truncates at second order for $z = \omega_1$, so that the $only$ contribution from $V$ is $P V P = \Omega \l(\ketbradt{1 C}{0 B} + \ketbradt{0 B}{1 C} \r)$ and the ground state shift $\frac{|\bra{0}T_S\ket{1}|^2}{\omega_1} \ketbrad{0 B}$. As seen in the supplemental section, $H_{eff} = \Sigma(\omega_1)$ is a good description of the evolution of states in $P$, and by our bounds on $\frac{|\bra{0}T_S\ket{1}|^2}{\omega_1} $ and $\delta \omega_1$, it causes the desired oscillations at a rate $2 \Omega$. 

Say that our fiducial state $\ket{F\, C}$ has overlap $|f_1| = |\!\bra{1} F \rangle|$. Since coherent oscillations between $\ket{1\,C}$ and $\ket{0 \, B}$ occur at frequency $2 \Omega$, as long as we evolve for times sampled randomly over a range $\frac{1}{\Omega}$, with probability $O(|f_1|^2)$ we expect to observe a transition of the form $\ket{1 \, C} \rightarrow \ket{0\, B}$, heralded by a measurment of the bath. The average simulation time of the algorithm then scales as $O\left(\frac{1}{|f_1|^2 \Omega} \right)$. If one is not given an explicit value of $\Omega$, as in \cite{Brassard2007} one may implement the scheme with evolution times sampled randomly from $[\tau,2 \tau]$, and iteratively increase $\tau \rightarrow 2 \tau$ after $\sim 1/|f_1|^2$ failed attempts. Since the sampling time $\tau$ grows exponentially, the total evolution time before success still scales as $O\left(\frac{1}{|f_1|^2 \Omega} \right)$.

Unfortunately, a bath measurement of $\ket{B}$ does {\it not} imply the system is in its ground state, as other resonant transitions could also occur. To account for this, after measuring $\ket{B}$ we map the bath state $\ket{B}\rightarrow \ket{L}$ and evolve under
\ba
\label{Hver}
H + \Omega_0 \mb{1}_S \otimes \left(\ketbradt{L}{R} + \ketbradt{R}{L} \right)
\ea
for time $\tau_v = \frac{\pi}{2 \Omega_0}$. The inversion symmetry of $H$ implies that only the system ground state $\ket{0}$ has degeneracy between its $\ket{L}$ and $\ket{R}$ bath states. All other eigenstates $\ket{\psi \, L}$ have at least $\Delta$ less energy than any $\ket{R}$ eigenstate. Hence by energy conservation only $\ket{0}$ exhibits coherent oscillations from $\ket{L}$ to $\ket{R}$, so measurement of the bath in $\ket{R}$ heralds success of the scheme. 

 Notice that, as long as $\delta \omega_1 \ll \Delta$, all of the constraints on $T_S$ required for this scheme are satisfied by $T_S = \Omega_0 \ketbrad{F}$. Since $\ketbrad{F}$ is rank 1, the assumption that the groundspace $P_0$ is non-degenerate is no longer necessary, as $T_S$ vanishes on every state in $P_0$ orthogonal to $f_0 \ket{0} = P_0 \ket{F}$ (where $|f_0|^2 = |\bra{F} P_0 \ket{F}|$). In this case we would have $\Omega \ket{1} = \Omega_0 f_0 P_1 \ket{F}$, so $|f_1|^2 = \bra{F}P_1 \ket{F}$ and $\Omega = |\Omega_0 f_0 f_1| $.  If we use unitary 2-designs to randomly generate $\ket{F}$, we require that there is a fixed probability in $n$ that $\Omega > \Omega_0 \sqrt{d_0\cdot d_1}/2^{n+1}$ and $|f_1|^2> d_1/2^{n+1}$, where $d_i$ is the rank of $P_i$. Thus if we are given no information about $H_S$ other than $\delta \omega_1$, $\omega_1$, and $\Delta$,  by using 2-designs and the probabilistic scheme we may obtain the ground state of $H_S$ with an average simulation time scaling as $O(\frac{2^{2 n}}{d_1^{3/2}d_0^{1/2}\Delta})$, reflecting the quadratic speedup observed in Section 2.

\section{Concluding Remarks}
There are several known alternatives to standard, logic-based quantum computing \cite{Farhi2000, Raussendorf2001, Mizel2001,Nayak2008, Childs2009,Verstraete2009,Nagaj2012}. The advantages of our scheme are that it requires the simulation of only time-independent Hamiltonians, as well as measurements of a single qubit (or qutrit) bath. Using the gadget construction  \cite{Jordan2008, Bravyi2008} we expect that it may be efficiently implemented with only 2-local interactions. 

Our work suggests several new avenues for investigation. One question is whether the techniques used in QSC may be applied to prepare other interesting states, such as mixed state ensembles \cite{Terhal2000,Temme2009,Ozols2012,Yung2012} or ground states of frustration free Hamiltonians \cite{Verstraete2009}. Using techniques in \cite{Magnus1954,Blanes2009,Davies1974}, one may attempt to show whether this scheme is robust against time dependent error terms in the simulation, or nonunitary evolution described by weak interactions with an environment. Finally, we note that the application of QSC to the clock Hamiltonian in Section 3 used only a 2-body system-bath interaction and a product fiducial state. Similarly to topological quantum computing, this prompts the question of when local interactions suffice to produce the ground state of a Hamiltonian, and fundamentally, what the relationship is between a Hamiltonian's computational complexity and the potential to cool it using such interactions.

\section*{Acknowledgements}
The authors would like to acknowledge Steven Jordan, Emanuel Knill, Yi-Kai Liu, Kristan Temme, and Frank Verstraete for valuable discussion. This work was funded by the NSF's Physics Frontier Center at the JQI.

\nointerlineskip \vspace{\baselineskip}
\noindent\hspace{\fill}\rule{1\linewidth}{2pt}\hspace{\fill}
\par\nointerlineskip \vspace{\baselineskip}

{\em  Supplemental Sections-}

Below we make rigorous the claims stated in the main body of the work. We start by developing some preliminary mathematical tools, then go on to give sufficient conditions for the success of the deterministic and probabilistic (extended) QSC schemes. We conclude by analysing the modified Kitaev clock Hamiltonian, thereby showing that both forms of QSC are polynomially equivalent to standard quantum computation.

\section{Mathematical Tools}
The following theorems ensure that the effective Hamiltonian, $H_{eff}$, that we derive accurately describes the dynamics of the simulator. The first result is a slight modification of Theorem 3 in \cite{Kempe2004}. The theorem is concerned with a Hamiltonian $H$ and a perturbation $V$ to the Hamiltonian. Within the subspace of interest, the theorem gives a one-to-one correspondence between the spectra of $H_{eff}$ and $\tilde H = H + V$, and bounds their difference. Before stating the theorem we require a few definitions.

Let $\mathbb{H}= P \oplus Q$ describe a finite dimensional Hilbert space on which $H$ acts, where $P$ is spanned by the eigenvectors of $H$ whose eigenvalues are in $(\lambda_-,\lambda_+)$ . Likewise define $\tilde P, \tilde Q$ with respect to $\tilde H$, using the same bounds. For simplicity we let the symbol for a subspace also represent its projector. We will assume that $H$ has gap $\Delta$, i.e. that the energies of $H$ in $P$ are at least $\Delta$ away from those in $Q$. We are interested in the dynamics under $\tilde H|_{\tilde P}$, which we approximate with $H_{eff}$. The approximation is derived from a series expansion of the self-energy operator \cite{Cohen-Tannoudji1992}:
\begin{flalign}
\Sigma_{P}(z) & = P H P + P V P +P V Q (z - Q(H + V)Q )^{-1} Q V P \nonumber \\
& = P H P + P V P + P V Q G(z) Q V P + P V Q G(z) Q V Q G (z) Q V P + ... \label{series}
\end{flalign}
In this notation $G(z) = (z - H)^{-1}$ is the Green's function for the unperturbed Hamiltonian.

In all cases below, $|| \cdot ||$ represents the operator 2-norm, 
$$||X|| = \sup_{\bra{v} v \rangle=1}  || X \ket{v}||$$
where $X$ is a (bounded) linear map from $\mbb{H}$ to a finite Hilbert space $\mbb{H}'$, and $|| \ket{v}||$ is the norm induced by the inner product of $\mbb{H}'$. This is a consistent norm, satisfying  $||A B||\leq ||A || \cdot ||B ||$ \cite{Stewart1990}. Finally, we mention a slight abuse of notation: If $\ket{v}$ is a vector in a Hilbert space and $\hat A$ an operator acting on that space, then for expressions of the form
\ban
\ket{v} + O(r) \\
\hat A + O(r)
\ean
$O(r)$ represents a vector (operator) with norm scaling as $O(r)$.  With this notation in hand, we can state the first theorem. Note that unless otherwise mentioned, the proofs for the following results are at the end of the section.
\\\\
\begin{theorem}[\cite{Kempe2004}]
\label{thm1}
 Assume that H has no eigenvalues in $[\lambda_- - \Delta/2,\lambda_- + \Delta/2]$ and $[\lambda_+-\Delta/2,\lambda_+ + \Delta/2]$, and that $||V||<\Delta/2$. Assume that there exists an operator $H_{eff}$ on $P$ whose spectrum is contained in $[c,d]$, and that for some $\gamma>0$, we have that
$$ [c - \gamma, d + \gamma] \subset (\lambda_-, \lambda_+)$$
$$ ||\Sigma_{P }(z) - H_{eff} || < \gamma$$
for all $z \in [c-\gamma,d+\gamma]$. Then if  $\lambda_j$ $(\tilde \lambda_j)$ is the $j$th largest eigenvalue of $H_{eff}$ $(\tilde H |_{\tilde P})$,  
$$|\tilde \lambda_j - \lambda_j|<\gamma$$
\end{theorem}

The result of \cite{Kempe2004} is for the case $\lambda_-= -\infty$. Since the proof of Theorem \ref{thm1} only requires a straightforward modification of the original, we do not show it here. The following result is used in the proof of Theorem \ref{thm1}, as well as in some of the claims below.\\

\begin{lemma}[\cite{Kempe2004}]
\label{lem1}
Let $H$, $\tilde H$ be two Hamiltonians with ordered eigenvalues $\mu_1 \leq \mu_2 \leq ...$ and $\sigma_1 \leq \sigma_2 \leq ...$ . Then for all $j$, 
$$|\mu_j - \sigma_j|\leq ||H - \tilde H||$$
\end{lemma}

Theorem \ref{thm1} gives bounds for the error in the approximate eigenvalues of $H_{eff}$, but in order to sufficiently describe the dynamics we also need a correspondence between the eigenvectors of $\tilde H$ and $H_{eff}$. To that end, we give a result derived Theorem 3.6, Chapter V, of Stewart and Sun's {\it Matrix Perturbation Theory} \cite{Stewart1990}. It is effectively a statement of the conservation of energy, and will ensure that transitions which do not preserve energy are suppressed. \\
\begin{theorem}[\cite{Stewart1990}]
\label{thm2}
Let $H$, $\tilde H$ be Hermitian operators.  Suppose $H$ is resolved by $P$ and $Q$: $H = P H P + Q H Q$, and
$$\textnormal{Spec}(H|_P) \subseteq [\lambda_- + \Delta/2, \lambda_+ - \Delta/2]\neq \emptyset $$
$$\textnormal{Spec}(H|_Q) \subseteq (-\infty, \lambda_- - \Delta/2] \cup [\lambda_++\Delta/2,\infty)$$
Let $\tilde P$ be the span of the eigenvectors of $\tilde H $ with eigenvalues contained in $(\lambda_-,\lambda_+)$.  Then for any $\ket{v}\in P$, $\ket{\tilde v} \in \tilde P$, 
$$\bra{\tilde v}  P \ket{\tilde v} \geq 1 -  \left(\frac{2 || H - \tilde H ||}{\Delta}\right)^2 $$
$$\bra{v} \tilde P \ket{v} \geq 1 -  \left(\frac{2 ||H - \tilde H ||}{\Delta}\right)^2$$
\end{theorem}
The proof of Theorem \ref{thm2} requires the following lemma. \\

\begin{lemma}
\label{lem2}
Let $A$, $B$ be Hermitian operators on $\mathbb{C}^{m},\mathbb{C}^{n}$, respectively. Suppose that $||A||\leq \alpha$  and that $B$ is invertible, with $||B^{-1}||\leq \frac{1}{\alpha + \beta}$, for $\beta>0$. Then for any linear operator $X : \mathbb{C}^n \rightarrow \mathbb{C}^m$,
\ban
||X|| \leq \frac{ || A X - X B||  }{\beta}
\ean
\end{lemma}

The proof of Lemma \ref{lem4} in the following section requires the following result. It will be used to bound the effect of error terms during time evolution:\\
\begin{lemma}
\label{lem3} Let $A$, $B$ be Hermitian operators on some space $\mb{S}_1\oplus \mb{S}_2$. Suppose that $A = \mb{S}_1 A \mb{S}_1 + \mb{S}_2 A \mb{S}_2 $, and that
\ban
&||\mb{S}_1 A^{-1} \mb{S}_1||& < 1/G_1\\
&||\mb{S}_2 A^{-1} \mb{S}_2|| &< 1/G_2\\
&b_{11} = ||\mb{S}_1 B \mb{S}_1||&<G_1/2\\
&b_{22}=||\mb{S}_2 B \mb{S}_2||&<G_2/2\\
&b_{12}=||\mb{S}_1 B \mb{S}_2||&<\min(G_1,G_2)/2
\ean
Then $(A-B)$ is invertible, and
\ban
&|| \mb{S}_1 (A - B)^{-1} \mb{S}_2 || & < \frac{b_{12}}{ (G_1 - b_{11})(G_2- b_{22}) - b_{12}^2}\\
&||\mb{S}_1 (A - B)^{-1} \mb{S}_1 || & < \frac{1}{G_1 - b_{11}}(1 + b_{12}|| \mb{S}_1 (A - B)^{-1} \mb{S}_2 ||) \\
&||\mb{S}_2 (A - B) ^{-1}\mb{S}_2  || & < \frac{1}{G_2 - b_{22}}(1 + b_{12} || \mb{S}_1 (A - B)^{-1} \mb{S}_2 ||)
\ean
 \end{lemma}

The goal of these theorems is to bound the error in the unitary evolutions designed to map $\ket{j \downn} \rightarrow \ket{0 \upp}$ or $\ket{1 \,C } \rightarrow \ket{0 \,B}$. This will be done by showing that both the eigenstates and eigenvalues corresponding to $H_{eff}$ are close to true eigenstates and eigenvalues of $\tilde H$.  Theorem \ref{thm1} is used to characterize the spectrum of $H_{eff}$. The following corrolary states that if an eigenspace $P'\subset P$ of $H_{eff}$ is 'well resolved' from its complement, then the corresponding eigenspace of $\tilde H$ is well approximated by $P'$.\\\\

\begin{corollary}
\label{cor1}
Given the assumptions of Theorem \ref{thm1}, let $P'\subset P$ be an eigenspace of $H_{eff}$, and let $ P - P'$ be its local complement. Define:
$$\nu =\max \left\{ |x - y| : x,y \in \textnormal{Spec}(H_{eff}|_{P'}) \right\}$$
$$ \eta = \min \left\{ |x - y| : x \in \textnormal{Spec}(H_{eff}|_{P'}), y \in \textnormal{Spec}(H_{eff}|_{P-P'}) \right\} $$
Define $\tilde P'$ as the eigenspace of $\tilde H$ obtained from the eigenvalue correspondence in Theorem \ref{thm1}. If $\eta > \gamma$, then for each eigenstate $\ket{\tilde v_i}\in \tilde P'$ of $\tilde H$, 
$$\bra{\tilde v_i} P'\ket{\tilde v_i} > \left( 1 - \left(\frac{2 ||V||}{\Delta} \right)^2 \right) \left(1 - \left(\frac{2 \gamma + \nu}{\eta - \gamma} \right)^2\right)$$
\end{corollary}
\begin{corollary}
\label{cor2}
Let $H$ be a Hamiltonian resolved by spaces $P$ and $Q$:  $H = P H P + Q HQ$. Assume that $P = P_1\oplus P_2 \oplus... \oplus P_L$, and that there exist $\lambda_{1-}<\lambda_{1+}<\lambda_{2-}<...<\lambda_{L+}$ such each $P_k$ corresponds to the eigenspace of $H$ with energies in $[\lambda_{k-}+\Delta/2, \lambda_{k+}-\Delta/2]$. Further assume that the eigenvalues of $H$ in $Q$ are at least $\Delta>0$ away from those in $P$. Finally, say that $\tilde P_k$ is the eigenspace of $\tilde H$ corresponding to eigenvalues within $(\lambda_{k-},\lambda_{k+})$, for $1\leq k \leq L$ 

Then for each $\ket{v}\in P$, $\ket{\tilde v} \in \tilde P = \tilde P_1 \oplus... \oplus\tilde P_L$,
\ban
\bra{v}\tilde P \ket{v} &\geq& 1 - L\left(\frac{2||\tilde H - H||}{\Delta}\right)^2\\
\bra{\tilde v}P \ket{\tilde v} &\geq& 1 - L\left(\frac{2||\tilde H - H||}{\Delta}\right)^2
\ean
\end{corollary}

\nointerlineskip \vspace{\baselineskip}
\noindent\hspace{\fill}\rule{1\linewidth}{.7pt}\hspace{\fill}
\par\nointerlineskip \vspace{\baselineskip}

\noindent We now begin the proofs of the above results, neglecting Theorem \ref{thm1} and Lemma \ref{lem1} as they may be derived (with minimal modification) from results in \cite{Kempe2004}.\\\\
\noindent{\it Proof of Lemma \ref{lem2}:}
Since $|| \cdot ||$ is a consistent norm, we have that 
$$|| A X || \leq \alpha ||X||$$
$$||X|| = ||X  B B^{-1} || \leq \frac{||X B||}{\alpha + \beta} \implies$$
$$ ||X B|| \geq (\alpha + \beta) ||X||$$
By the triangle inequality we conclude that
\ban
||A X - X B|| \geq ||X B|| - ||A X|| \geq  \beta ||X||
\ean
\begin{flushright}
$\square$
\end{flushright}

\noindent{\it Proof of Lemma \ref{lem3}:}\\
First, we show that $(A- B)$ is invertible. It suffices to show that $||B A^{-1}||<1$, as then $(1 - B A^{-1}) = \left((A - B) A^{-1}\right)$ is invertible. Let a normalized $\ket{v_i}\in \mb{S}_i$ be given. Then $A^{-1} \ket{v_i} = c \ket{v_i'}$ for some (normalized) $\ket{v_i'}\in \mb{S}_i$ and $|c|<1/G_i$. Using the above equation, one gets
\ban
||B A^{-1} \ket{v_i}||^2 & = & \bra{v_i}  A^{-1} B  B A^{-1} \ket{v_i}\\
& < & \frac{1}{G_i^2} \bra{v_i'}  B \left( \mb{S}_1 + \mb{S}_2\right) B  \ket{v_i'}\\
& \leq & \frac{b_{i i}^2 + b_{12}^2}{G_i^2}\\
& < & 1/2
\ean
For any normalized state $\ket{v}$, one may write $\ket{v} = a \ket{v_1} + b \ket{v_2}$, with $\ket{v_i}\in \mb{S}_i$ and $|a|^2 + |b|^2 = 1$, and by the triangle inequality
\ban
||B A^{-1} \ket{v} || &\leq& |a| \cdot||B A^{-1} \ket{v_1}|| + |b|\cdot  ||B A^{-1} \ket{v_2}||\\
& < & \frac{1}{\sqrt{2}}( |a| + |b| ) \leq 1
\ean
where the last inequality follows from writing $|a| = \sqrt{s}, |b| = \sqrt{1-s}$ for some $s \in [0,1]$. This shows that $||B A^{-1}||<1$, which implies that $(1 - B A^{-1})$ has only positive eigenvalues and is therefore invertible.
 
Since we have shown that $(A-B)$ is invertible, one may easily check that
\ba
\label{geom}
(A - B)^{-1} = A^{-1} +  A^{-1} B  (A - B)^{-1} 
\ea
Decomposing into $\mb{S}_1$ and $\mb{S}_2$ components, one gets
\ban
\mb{S}_1(A - B)^{-1}\mb{S}_1 & = & \mb{S}_1 A^{-1} \mb{S}_1  +  \mb{S}_1 A^{-1} \mb{S}_1 \left( \mb{S}_1 B \mb{S}_1 (A-B)^{-1} \mb{S}_1 + \mb{S}_1 B \mb{S}_2 (A-B)^{-1} \mb{S}_1 \right)
\ean
By the triangle inequality and the relation $|| CD || \leq || C||\cdot  || D ||$, we have
\ba
\nonumber || \mb{S}_1(A - B)^{-1}\mb{S}_1 || &<& \frac{1}{G_1}  + \frac{b_{11}}{G_1} || \mb{S}_1(A - B)^{-1}\mb{S}_1 || + \frac{b_{12}}{G_1} || \mb{S}_1(A - B)^{-1}\mb{S}_2 || \implies \\
|| \mb{S}_1(A - B)^{-1}\mb{S}_1 || & < & \frac{1}{G_1 - b_{11}}\left(1 + b_{12} || \mb{S}_1(A - B)^{-1}\mb{S}_2 ||  \right)\label{SSbound}
\ea
where we also used the fact that we may interchange the projectors $\mb{S}_1,\mb{S}_2$ when taking the operator norm. Interchanging the numbers 1 and 2, we obtain an equivalent result for $ ||\mb{S}_2(A - B)^{-1}\mb{S}_2||$. This proves the last two statements in Lemma \ref{lem3}. 

Likewise, using \eqref{geom} one may show that 
\ba
\nonumber || \mb{S}_1(A - B)^{-1}\mb{S}_2 || & = & ||\mb{S}_1 A^{-1}\mb{S}_1 \left(  \mb{S}_1 B \mb{S}_1 (A - B)^{-1} \mb{S}_2 + \mb{S}_1 B \mb{S}_2 (A - B)^{-1} \mb{S}_2 \right)  || \\ 
\nonumber & \leq & \frac{b_{11}}{G_1} || \mb{S}_1(A - B)^{-1}\mb{S}_2 || + \frac{b_{12}}{G_1}||  \mb{S}_2 (A - B)^{-1} \mb{S}_2  || \implies \\
|| \mb{S}_1(A - B)^{-1}\mb{S}_2 || & \leq &  \frac{b_{12}}{G_1-b_{11}}  ||  \mb{S}_2 (A - B)^{-1} \mb{S}_2  || 
\ea
Substituting the result of \eqref{SSbound} (with $1$ and $2$ interchanged) into the right hand side produces the first statement of Lemma \ref{lem3}.
\begin{flushright}
$\square$
\end{flushright}

\noindent{\it Proof of Theorem \ref{thm2}:}\\

Define the numbers 
$$\bar \lambda = \frac{\lambda_+ + \lambda_-}{2}$$
$$\Lambda = \frac{\lambda_+ - \lambda_-}{2}$$
Note that $\Lambda\geq\Delta/2$ since $(\lambda_+ - \Delta/2)\geq (\lambda_-+\Delta/2)$. Since $H$ is Hermitian, there exists a unitary operator $U = ( X_ P,  X_Q)$ that diagonalizes $H$, where the columns of $X_P$ and $X_Q$ form an orthonormal basis for $P$ and $Q$, respectively. We see that
$$H- \bar \lambda \mb{1} = X_P L_P X_P^\dagger +  X_Q L_Q X_Q^\dagger$$
where $L_p$ and $L_Q$ are diagonal matrices, with eigenvalues in $[-(\Lambda -\Delta/2),\Lambda - \Delta/2]$ and $(-\infty,- \Lambda-\Delta/2]\cup [\Lambda + \Delta/2,\infty)$, respectively. Analogously, we may define the decomposition of $\tilde H - \bar \lambda$:
$$\tilde H- \bar \lambda \mb{1} = \tilde X_{\tilde P} \tilde L_{\tilde P} \tilde X_{\tilde P}^\dagger +  \tilde X_{\tilde Q} \tilde L_{\tilde Q} \tilde X_{\tilde Q}^\dagger$$
where the eigenvalues of  $\tilde L_{\tilde P}$ and $ \tilde L_{\tilde Q}$ are in $[-\Lambda,\Lambda]$ and $(-\infty, -\Lambda]\cup [\Lambda ,\infty)$. 

Consider the operator
$$V_E = \tilde X_{\tilde P}^\dagger \left( \tilde H - H  \right) X_Q$$
From the identities above it follows that
\ban 
V_E & = & \tilde X_{\tilde P}^\dagger ( \tilde H - \bar\lambda \mb{1})  X_Q - \tilde X_{\tilde P}^\dagger (  H - \bar\lambda \mb{1})X_Q \\
& = &   \tilde L_{\tilde P} \tilde X_{\tilde P}^\dagger X_Q -  \tilde X_{\tilde P}^\dagger X_Q L_Q 
\ean
Noting that $||\tilde L_{\tilde P}||\leq \Lambda$ and $||L_Q^{-1}||\leq \frac{1}{\Lambda + \Delta/2}$, we may use Lemma \ref{lem2} with $A = \tilde L_{\tilde P}$, $B = L_Q$, $\alpha = \Lambda$, $\beta = \Delta/2$, and $X = \tilde X_{\tilde P}^\dagger X_Q$ to conclude
\ban
||\tilde X_{\tilde P}^\dagger X_Q||& \leq & \frac{2 ||V_E|| }{\Delta} \\
& = & \frac{2 ||\tilde X_{\tilde P}^\dagger \left( \tilde H - H  \right) X_Q|| }{\Delta}\\ 
& \leq & \frac{2 ||\tilde H - H  || }{\Delta}
\ean
Where the last line follows from $||A B ||\leq ||A ||\cdot  || B ||$. 

Let $\ket{\tilde v} \in \tilde P$ be given with $\bra{\tilde v} \tilde v \rangle = 1$. Since the columns of $\tilde X_{\tilde P}$ form an orthonormal basis for $\tilde P$, there exists a vector $\ket{x}$ such that $\ket{\tilde v} = \tilde X_{\tilde P} \ket{x}$ and $\bra{x}\! x \rangle = 1$. From the previous line we conclude that
\ban
\bra{\tilde v} Q \ket{\tilde v} & = & \bra{\tilde v} X_Q X_Q^\dagger \ket{\tilde v}\\
& = &  \bra{ x} \tilde X_{\tilde P}^{\dagger} X_Q X_Q^\dagger \tilde X_{\tilde P} \ket{x}\\
& \leq & || X_Q^\dagger \tilde X_{\tilde P}||^2 \\
& = & ||\tilde X_{\tilde P}^\dagger X_Q||^2 \leq \left(\frac{2 ||\tilde H - H  || }{\Delta}\right)^2
\ean
where in the last line we used the fact that the operator 2-norm is unchanged when taking the Hermitian adjoint.
The first statement follows by noting that $Q = \mb{1} - P$. To prove the second statement, we can make an identical argument using $\tilde V_E = X_P^\dagger (\tilde H - H) \tilde X_{\tilde Q}$.
\begin{flushright}
$\square$
\end{flushright}

\noindent{\it Proof of Corollary \ref{cor1}:}

By assumption, we have that $H = P H P + Q H Q$, where $P$ projects into the eigenspace of $H$ with energy in $(\lambda_- + \Delta/2, \lambda_+ - \Delta/2)$. This defines the energy gap between $H|_P$ and $H|_Q$. For $\tilde H = H + V$, we define $\tilde P$ as the eigenspace of $\tilde H$ with energy in $(\lambda_- ,\lambda_+)$. $P'\subset P$ is an eigenspace of $H_{eff}$ and $\tilde P'\subset \tilde P$ is the associated eigenspace of $\tilde H$, obtained from the eigenvalue correspondence discussed in Theorem \ref{thm1}. 

Let $\ket{\tilde v_i} \in \tilde P'$ be an eigenstate of $\tilde H$, with corresponding energy $\tilde E_i$.  We may write
$$\ket{\tilde v_i} = N_i \ket{u_i} + Q \ket{\tilde v_i}$$
where $P \ket{\tilde v_i} = N_i \ket{u_i}$. Since $\ket{\tilde v_i} \in \tilde P$, we may use Theorem \ref{thm2} to conclude that
$$N_i^2 = \bra{\tilde v_i} P \ket{\tilde v_i} \geq 1 - \left(\frac{2 ||V||}{\Delta} \right)^2$$ 

Now decompose $\ket{u_i}$ into components parallel and orthogonal to $P'$:
$$\ket{u_i} = P' \ket{u_i} + (P-P')\ket{u_i}$$
We wish to bound $ \bra{u_i} P' \ket{u_i}$ from below. To do this we first note that, as in the proofs of Lemmas 5 and 6 of \cite{Kempe2004}, $\Sigma_P(\tilde E_i) \ket{u_i} = \tilde E_i \ket{u_i}$. From this we conclude that
$$ (\Sigma_P(\tilde E_i) - H_{eff}) \ket{u_i} + (H_{eff} -\tilde E_i \mb{1} )P'\ket{u_i} = -( H_{eff} -\tilde E_i \mb{1})(P-P')\ket{u_i}$$
By assumption, $|| (\Sigma_P(\tilde E_i) - H_{eff})||<\gamma$. Since $E_i$ is an eigenvalue corresponding to $P'$, the eigenvalues of operator $(H_{eff}-E_i \mb{1})P'$ have magnitude at most $\nu$, and since $|E_i - \tilde E_i|<\gamma$, $||(H_{eff} - \tilde E_i \mb{1}) P' ||<\gamma + \nu$. By the triangle inequality, the norm of the left hand side is less than $2 \gamma + \nu$. Likewise, the eigenvalues of $(H_{eff} - E_i \mb{1})|_{P-P'}$ have magnitude at least $\eta$ so the eigenvalues of $(H_{eff} - \tilde E_i \mb{1})|_{P-P'}$ are greater than $\eta - \gamma >0$. Thus the norm of the right hand side is greater than $(\eta - \gamma) ||(P-P')\ket{u_i}||$. Therefore
$$ ||(P-P')\ket{u_i}|| < \frac{2 \gamma + \nu}{\eta - \gamma} \implies$$
$$ \bra{u_i} P' \ket{u_i} = 1 -  \bra{u_i} (P-P') \ket{u_i} > 1 - \left(\frac{2 \gamma + \nu}{\eta - \gamma} \right)^2$$
The result follows by noting that $\bra{\tilde v_i} P'\ket{\tilde v_i} = N_i^2  \bra{u_i} P' \ket{u_i}$.
\begin{flushright}
$\square$
\end{flushright}

\noindent{\it Proof of Corollary \ref{cor2}:}\\
We may assume without loss of generality that each domain $[\lambda_{k-}+\Delta/2,\lambda_{k+}-\Delta/2]$ is at least $2 \Delta$ away from its neighboring domains. If this were not the case for a neighboring domain, the gap $\Delta$ from $\mbox{Spec}(H|_{Q})$ would imply that there exist no $QHQ$ eigenvalues between the two domains, so they may be merged.

Let a normalized vector $\ket{v}\in P$ be given. We may decompose $\ket{v}$ into its components within each $P_k$:
$$\ket{v} = \sum_{k}c_k \ket{v_k}$$
where $\sum_k |c_k|^2 = 1$.  We may apply Theorem \ref{thm2} to any given $P_k$ to get 
$$ 1 - |r_k|^2 = \bra{v_k}\tilde P \ket{v_k}\geq \bra{v_k}\tilde P_k \ket{v_k}$$
where $|r_k| \leq \frac{2 ||\tilde H - H||}{\Delta}$. So for each $k$,
$$\ket{v_k} = \tilde P \ket{v_k} + r_k \ket{v_k^\perp}$$
where
$$\ket{v_k^\perp} \in \tilde Q$$

Substituting into the definition of $\ket{v}$, we get
\ban
\ket{v} &=& \sum_{k}c_k \left(\tilde P \ket{v_k} + r_k \ket{v_k^\perp} \right)\\
& = & \tilde P \left( \sum_{k}c_k \ket{v_k}\right) +  r_{eff} \ket{v^\perp}
\ean
where $r_{eff} \ket{v^\perp} = \sum_k c_k r_k \ket{v_k^\perp}$ is a vector in $\tilde Q$. By the triangle and Cauchy-Schwartz inequalities we conclude that 
\ban
|r_{eff}|&\leq& \sum_{k} |c_k r_k| \\
&\leq& \sum_k |c_k| \l(\frac{2 ||\tilde H - H||}{\Delta}\r) \\
 &\leq& \sqrt{L}\left(\frac{2 ||\tilde H - H||}{\Delta}\right)
\ean
The first statement of the corollary follows by noting that $\bra{v} \tilde P \ket{v} = 1 - |r_{eff}|^2$. The second statement follows by making the same argument and, except with the initial assumption, adding or removing $\sim$ to each projector and vector.
\begin{flushright}
$\square$
\end{flushright}

\section{Deterministic QSC Analysis}
This section describes the deterministic Quantum Simulated Cooling scheme, and gives sufficient conditions for its success up to an infidelity $O(\epsilon)$.\\
\begin{theorem}
\label{thm3}
Let $H_S$ be a Hamiltonian acting on $n$ qubits, with eigenspaces $\mb{S}_1$ and $\mb{S}_2$. Assume that 
$$\mb{S}_1 H_S \mb{S}_1= \sum_{j=0}^{L} \omega_j \ketbrad{j}$$
where $0 = \omega_0 < \omega_1 < \omega_2 ... < \omega_L$ and
$$\Delta = \min_{\omega_i,E}\{ |E - \omega_i| : E \in \textnormal{Span}(H_S), E\neq \omega_i \} > 0$$
Finally, assume that there exists operators $T_S$, $\hat \delta_j$, such that 
$$\mb{S}_1 T_S \mb{S}_1 = \Omega_0 \ketbrad{G}$$
with 
$$\ket{G} = \sum_{j=0}^L x_j \ket{j}$$
$|x_j|>0$, and for all $j$,
\ba
\label{errorbound}
\begin{split}
 r = \Omega_0 / \Delta& < 1/8\\
\frac{||\mb{S}_1 \hat \delta_j \mb{S}_1||}{\Delta}& < r\cdot \Omega_0 |x_0 x_j|/\Delta\\
\frac{||\mb{S}_1 (T_S + \hat \delta_j)\mb{S}_2||^2}{\Delta^2} &< r \cdot  \Omega_0 |x_0  x_j|/\Delta \\
\frac{||\mb{S}_2 (T_S + \hat \delta_j) \mb{S}_2 || }{\Delta}&< 1/2
\end{split}
\ea
Then for $\epsilon>0$ and  $r \propto \epsilon L^{-5/2}$, there exists a TCP map $\mbb{E}$, consisting of $L$ unitaries $U_j$ on $n+1$ qubits of the form 
$$U_j = \exp\left(-i \tau_j (H_S + \omega_B^{(j)}\ketbrad{\upp}^{(n+1)} + T_S \otimes \sigma_x^{(n+1)} + \hat \delta_j)\right)$$
and single qubit measurements, such that for any state $\ket{F}\in \mb{S}_1$,
$$\textnormal{Tr}[ (\ketbrad{0}\otimes \mb{1}_{n+1}) \mbb{E}( \ketbrads{F \downn}) ] = 1 - O(\epsilon)$$
as $\epsilon\rightarrow 0^+$, with total simulation time
\ban
T = O\left(\frac{L^{5/2}}{\epsilon \Delta |x_0|}\right) \sum_{j = 1}^L \frac{1}{|x_j|}
\ean
\end{theorem}
{\it Proof:}

The proof is constructive. $\mbb{E}$ can be described by a loop of $L$ cooling steps, labeled by the index $j$ (starting at $j = L$). At the beginning of each iteration, the bath is measured in the logical basis. If it is in state $\ket{\upp}$, then the transition to the ground state has already occured, so we effectively terminate by decrementing $j \rightarrow j-1$ and continuing to the next iteration. Otherwise, we evolve under Hamiltonian $\tilde H_{j} = H_j + V + \hat \delta_j$, where $H_j = H_S + \omega_B^{(j)} \mb{1}_S\otimes \ketbrads{\upp}_B$, $V = T_S \otimes \sigma_x$, and $\hat \delta_j$ represents an error term in the simulation. The bath energy $\omega_B^{(j)}$ (defined explicitly below) is near $\omega_j$. The time evolved under this Hamiltonian is $\tau_j = \frac{\pi}{2 \Omega_0 |x_0 x_j|}(1 + O(r^2))$, and the associated unitary evolution is labeled $U_j$. We then decrement $j \rightarrow j-1$ and continue to the next iteration. 

Written in pseudocode, $\mbb{E}$ may be summarized as
\begin{verbatim}
For j= L, L-1, ... 1
     Measure bath qubit
     If bath is down:
          Apply U_j  
Measure bath
\end{verbatim}
Step $j$ is associated with the following trace preserving, completely positive (TCP) map:
\ba
\begin{split}
E_j(\rho) = U_j \ketbradts{\!\downn}{\downn} &  \rho  \ketbradts{\!\downn}{\downn} U_j^\dagger \\
 + \ketbradts{\!\upp}{\upp} &  \rho  \ketbradts{\!\upp}{\upp}\label{Ej}
\end{split}
\ea
where $\ketbradts{\!\!\downn}{\downn\!}$ represents the operator $\left(\mb{1}_S \otimes \ketbrads{\downn}_B \right)$, and likewise for $\ketbradts{\!\!\upp}{\upp\!}$. We define $\mathbb{E} =  E_1 \circ E_2 ... \circ E_L$. Given the above assumptions, the lemmas below prove that the algorithm works as expected. The proofs of the lemmas are included at the end of this section.\\\\
\begin{lemma}[Fidelity of the cooling step]
\label{lem4}
Let $U_j = \exp(-i \tau_j \tilde H_j )$ be the unitary evolution associated with cooling state $\ket{j \downn}$. Assume that \eqref{errorbound} holds. There exists a time $\tau_j = \frac{\pi}{2 \Omega_0 |x_0x_j|}(1 + O(r^2))$ and $\omega_B^{(j)} \in (\omega_j-\Delta/4,\omega_j + \Delta/4)$ such that
$$U_j \ket{j \downn}=  \ket{0 \upp} + O(r)$$
where the bound $O(r)$ is uniform over all $j$.
\end{lemma}\\
\begin{lemma}[Preservation of the lower bands]
\label{lem5}
Define
$$ M_{j-1}^d = \textnormal{Span}\{\ket{k \downn} \}_{k=0}^{j-1}$$
Then under the assumptions of Lemma \ref{lem4}, for any state $\ket{v } \in M_{j-1}^d$,
$$ U_j \ket{v} = M_{j-1}^d U_{j} \ket{v} + \sqrt{j}\cdot O(r)$$
\end{lemma}
\begin{lemma}[Projective mapping of TCP map]
\label{lem6}
Define 
\ban
M_j & =  &M_{j}^d\oplus \textnormal{Span}\left(\{\ket{0 \upp} \}\right)
\ean 
Then for $M_j \rho M_j = \rho$, 
$$E_j(\rho) = \lambda_j \rho_{j-1} + \hat R_j $$
where $\rho_{j-1}$ is a density matrix satisfying $M_{j-1}\rho_{j-1} M_{j-1} = \rho_{j-1}$, $\lambda_j = 1 - \textnormal{Tr}[\hat R_j]$, $\textnormal{Tr}[\hat R_j] = j^{3/2}\cdot O(r)$, and $||\hat R_j|| = \sqrt{j} \cdot O(r)$. 
\end{lemma}
\\
\begin{lemma}[Success of the Deterministic Algorithm]
\label{lem7}
Given the assumptions in the previous lemmas, let $\rho = \ketbradts{F \!\!\downn}{F\!\!\downn\!}$ for some $\ket{F } \in \mb{S}_1$, and assume that $r = \Omega_0/\Delta \propto \epsilon L^{-5/2}$. Then
$$ \textnormal{Tr}[ M_0 \mathbb{E}(\rho)] = 1 - O(\epsilon)$$
as $\epsilon \rightarrow 0$.
\end{lemma}
\\\\
\indent Since $M_0 = \ketbrad{0}\otimes \mb{1}_{n+1}$, Lemma \ref{lem7} proves the first claim in Theorem \ref{thm3}. As shown in the proof of Lemma \ref{lem4}, unitary $U_j$ requires an evolution time $\tau_j = \frac{\pi}{2 \Omega}$, where $\Omega = \Omega_0 |x_0 x_j|(1 + O(r))$. The simulated time required for the algorithm is therefore
\ban
T &= & \sum_{j =1}^L \tau_j \\
& = &  \frac{\pi}{2 \Omega_0 |x_0| }\sum_{j = 1}^L \frac{1}{|x_j|}(1 + O(r))\\
& = & O\left(\frac{L^{5/2}}{\epsilon \Delta |x_0|}\right) \sum_{j = 1}^L \frac{1}{|x_j|}
\ean
where we used the fact that $\Omega_0^{-1} = (r \Delta)^{-1} = \Delta^{-1} L^{5/2}\cdot O(\epsilon^{-1})$.
\begin{flushright}
$\square$
\end{flushright}
\subsection{Reduced Algorithm:}

It is possible that the decomposition $\ket{G} = \sum_{j} x_j \ket{j}$ contains values of $x_j$ that are exponentially small in $n$, meaning that the simulation time $\tau_j = O(\exp(n))$. We therefore discuss conditions under which it is valid to neglect the cooling of such states, leading to an improved run time. In doing so we derive bounds on the allowed error in the preparation of the fiducial state $\ket{F \downn}$.

Suppose that we choose to skip the cooling of states $\ket{j}$ such that $|x_j| \leq \eta$, for some parameter $\eta$. Defining $\mb{S}_1' = \mbox{Span}\{ \ket{j} : |x_j|>\eta \} \otimes \mathbb{H}_B$, we may write
$$\ket{F \downn} = \mb{S}_1' \ket{F \downn} + f_\perp \ket{F_\perp \downn}$$ 
Given $\rho = \ketbradt{F \downn}{F \downn \!}$, we may compute
\ba
\rho = (1 - |f_\perp|^2) \ketbradt{F' \downn}{F' \downn\!} + \hat R_\perp
\ea
where $\ket{F' \downn}\in \mb{S}_1'$, and $\hat R_\perp$ is an error term with rank $2$ and norm $||\hat R_\perp||\leq \frac{3}{\sqrt{2}} |f_\perp|$.

Defining $\mb{S}_1^\perp$ as the complement of $\mb{S}_1'$ in $\mb{S}_1$, we can write $\mb{S}_2' = \mb{S}_2 + \mb{S}_1^\perp$. From \eqref{errorbound} and the triangle inequality,
\ba
\begin{split}
||\mb{S}_1'(V + \hat \delta)\mb{S}_2'|| &= O(\Delta r)\\
||\mb{S}_2'( \hat \delta)\mb{S}_2'||& = \Delta( 1/2 + O(r))
\end{split}
\ea
As these are the only assumptions necessary to prove Lemma \ref{lem5}, for sufficiently small $r$ it still holds for the reduced space $(M^d_{j-1})'  = M_{j-1}^d\cap\mb{S}_1'$. Combining this with Lemma \ref{lem4}, we see (as in its proof) that Lemma \ref{lem6} also holds with respect to the space $M_j' = M_j \cap \mb{S}_1'$. We may then apply Lemma \ref{lem7} to the density matrix $\rho' = \ketbrads{F' \downn}$ and get the same fidelity $\epsilon$ for the reduced algorithm $\mathbb{E}' = E_{i_1}\circ ... \circ E_{i_{L'}}$, where $i_1 \leq i_2 ... \leq i_{L'}$ enumerate the eigenstates in $\mb{S}_1'$. 

Since $\mathbb{E}'$ is applied to $\rho$ and not $\rho'$, we see that the reduced algorithm fidelity is 
\ba
\begin{split}
\mbox{Tr}[M_0 \mathbb{E}'(\rho)] & = &(1 - |f_\perp|^2)\mbox{Tr}[M_0 \mathbb{E}'(\rho')] + \mbox{Tr}[M_0 \mathbb{E}'(\hat R_\perp)] \\
& = & (1 - |f_\perp|^2)(1 - O(\epsilon)) + \mbox{Tr}[M_0 \mathbb{E}'(\hat R_\perp)]
\label{fidelity2}
\end{split}
\ea
Because $\mbb{E}'$ is a composition of $L'$ single qubit measurements and unitary evolutions, as seen in the proof of Lemma \ref{lem7} we have that $|| \mathbb{E}'(\hat R_\perp)|| \leq L' \cdot ||\hat R_\perp|| = L' \cdot O(|f_\perp|)$, where $L'=\mbox{dim}(\mb{S}_1')/2 - 1\leq L$. Furthermore, as $M_0$ is a rank 2 projector, $M_0 \mathbb{E}'(\hat R_\perp)$ has rank at most $2$, so we conclude that 
\ban |\mbox{Tr}[M_0 \mathbb{E}'(\hat R_\perp)]| \leq \mbox{rank}(M_0 \mathbb{E}'(\hat R_\perp))\cdot || M_0 \mathbb{E}'(\hat R_\perp)|| = L' \cdot O(|f_\perp|) \ean
Using this result in \eqref{fidelity2}, as long as $|f_\perp| = O(\epsilon/L')$, we may still achieve fidelity $1 - O(\epsilon)$. Notice this allows us to treat any error in the preparation of $\ket{F \downn}$ as contributing to $|f_\perp|$.

Since we are only cooling states such that $|x_j|>\eta$, the timing of the algorithm is then
\ban
T' &=& O \left(\frac{1}{\Omega_0 |x_0|} \sum_{k = 1}^{L'} \frac{1}{|x_{i_k}|}\right)\\
& \leq & O \left(\frac{1}{ \Omega_0 |x_0|} \frac{ L'}{\eta}\right)\\
& = & O \left(\frac{(L')^{7/2}}{\epsilon \eta \Delta |x_0|} \right)
\ean
In the case where $\ket{F} = \ket{G}$ we see that $|f_\perp|^2 = \sum_{|x_j|\leq\eta} |x_j|^2 \leq L \eta^2$, so $L \cdot |f_\perp| \leq L^{3/2}\eta$ and in order to maintain an infidelity $O(\epsilon)$ it is sufficient to set $\eta = \epsilon/L^{3/2}$. This gives 
\ban
T' = O \left(\frac{L^5}{\epsilon^2 \Delta |x_0|} \right)
\ean
\subsection{Proofs of Lemmas \ref{lem4}-\ref{lem7}}

The proof of Lemma \ref{lem4} is the most involved. First, we analyze $\tilde H_j$ in the subspace $\mb{S}_1\otimes \mbb{H}_B$, and show that it has eigenstates near $\ket{j \downn} \pm \ket{0 \upp}$, with energies near $\omega_B \pm \Omega_0 x_0 x_j$. In order to do this we compute the self energy operator $\Sigma_P(z)$ for the manifold of energies near $\omega_j$, and show how the value of $\omega_B^{(j)}$ may be calculated to account for energy shifts associated with $V$.  We then account for the static error term $\hat \delta$, and the possibility that $V$ may couple between $\mb{S}_1$ and $\mb{S}_2$.
\\\\\noindent {\it Proof of Lemma \ref{lem4}:}

We begin by analyzing the case where $\hat \delta = 0$ and $V = \mb{S}_1 V \mb{S}_1 = \Omega_0 \ketbrad{G}\otimes \sigma_x$. Note that $\mb{S}_1 = \left(\mbox{Span}\{\ket{k} \}_{k=0}^L\right)\otimes \mathbb{H}_B$ is then invariant under $\tilde H_j = H_S + \omega_B^{(j)}\ketbrad{\upp}^{(n+1)} + T_S\otimes \sigma_x$, which will simplify our analysis. In order to understand the dynamics of $ \tilde H_j $, we compute the level shift operator \eqref{series} to find an effective Hamiltonian for the eigenstates of $\tilde H_j$ with energy near $\omega_j$. Let $P$ be the eigenspace of energies between $\lambda_- = \omega_j - \Delta/2$ and $\lambda_+ = \omega_j + \Delta/2$. Given that the eigenstates of $H_S$ in $\mb{S}_2$ are at least $\Delta$ away from those in $\mb{S}_1$, and $|\omega_B^{(j)}-\omega_j|<\Delta/4$, we have
\ban
P &=& \ketbradt{ j \downn}{ j \downn\!}+ \ketbradt{0 \upp}{0 \upp\!}\subset \mb{S}_1\\
Q & = & \sum_{k \neq 0} \ketbradt{k \upp}{k \upp\!} + \sum_{k \neq j} \ketbradt{k \downn}{k \downn \!} + \mb{S}_2 Q \mb{S}_2 \\
P H_j P & = & \omega_j \ketbradt{ j \downn}{ j \downn\!}+ \omega_B^{(j)} \ketbradt{0 \upp}{0 \upp\!}\\
 V & = & \Omega_0 \ketbrad{G}\otimes \sigma_x
\ean

Using \eqref{series}, we may now compute 
\ba
\begin{split}
\label{sigmaz}
\Sigma_P(z)  &= P H_j P + P V P  + P V \frac{Q}{z - Q(H_j + V)Q} V P  \\
 &=  P H_j P + P V P + P V \left(G_Q(z) +  G_Q(z) V G_Q(z)  + ... \right) V P 
\end{split}
\ea
where $G_Q(z) = \frac{Q}{z - Q H_j Q}$. Since the projector into $\mb{S}_1$ commutes with $P,Q, H_j$, and $V$, we may replace all operators by their projections into $\mb{S}_1$ above. This simplifies $\Sigma_P(z)$, giving
\ba
\begin{split}
\label{sigmaz2form}
 \Sigma_P(z) & =  \omega_j \ketbradt{ j \downn}{ j \downn\!}+ \omega_B^{(j)} \ketbradt{0 \upp}{0 \upp\!}\\
& \quad  + \Omega_0 P \ket{G} \left(\sigma_x +\Omega_0 \sigma_x \bra{G}G_Q(z) \ket{F} \sigma_x\right) \sigma_*(z)\bra{F} P
\end{split}
\ea
where $\sigma_*$ is the bath operator
\ban
\sigma_*(z) = (1 + \left(\Omega_0 \bra{G}G_Q(z) \ket{G} \sigma_x\right)^2 + \left(\Omega_0 \bra{G}G_Q(z) \ket{G} \sigma_x\right)^4+ ...)
\ean
As a bath operator, one has that
$$ \bra{G}G_Q(z) \ket{G} = g_j(z) \ketbrad{\downn} + g_0(z) \ketbrad{\upp}$$
$$  g_j(z) = \bra{G \downn} G_Q(z)\ket{G \downn} = \sum_{k\neq j} \frac{|x_k|^2}{z - \omega_k}$$
$$  g_0(z) = \bra{G \upp} G_Q(z)\ket{G\upp}= \sum_{k\neq 0} \frac{|x_k|^2}{z - (\omega_k + \omega_B^{(j)})}$$
so 
$$\left(\Omega_0 \bra{G}G_Q \ket{G} \sigma_x\right)^{2n} = \left(\Omega_0^2 g_jg_0\right)^n (\ketbrad{\downn} + \ketbrad{\upp})$$ 
therefore
\ba
\sigma_*(z) = \frac{1}{1 - \Omega_0^2 g_j(z)g_0(z)} (\ketbrad{\downn} + \ketbrad{\upp})
\label{sigmastar}
\ea
Since $P \ket{G \downn} =  x_j \ket{j \downn}$, $P \ket{G \upp} = x_0 \ket{0 \upp}$, we may use \eqref{sigmaz2form} and \eqref{sigmastar} to get
\ba
\label{Heff1}
\begin{split}
 \Sigma_P(z)  &=  \omega_j \ketbradt{ j \downn}{ j \downn\!}+ \omega_B^{(j)} \ketbradt{0 \upp}{0 \upp\!}\\
& \quad  +\frac{ \Omega_0}{1 - \Omega_0^2 g_jg_0} P \Big(\ketbradt{G \downn}{G\upp}  + \ketbradt{G\upp}{G \downn}  \\
& \quad  + \Omega_0 g_j\ketbrad{G \upp} +\Omega_0 g_0\ketbrad{G \downn} \Big) P \\
& =   \left(\begin{array}{c c}
\omega_j & 0\\
0 & \omega_B^{(j)} 
\end{array} \right) +\frac{\Omega_0}{1 - \Omega_0^2 g_j(z) g_0(z)} \left(\begin{array}{c c}
|x_j|^2 \Omega_0 g_0(z)  & x_0 x_j\\
x_0 x_j & |x_0|^2 \Omega_0 g_j(z)
\end{array} \right)
\end{split}
\ea
where the above matrices are written in the $\{\ket{j \downn}, \ket{0 \upp}\}$ basis. 

Using the above expression, we define the effective Hamiltonian $ H_{eff} = \Sigma_P(\omega_B^{(j)})  $. We identify the diagonal elements of the second matrix above as the level shift, and observe these are composed of even powers of $\Omega_0$. If the diagonal terms in $H_{eff}$ were equal, it would cause coherent oscillations $\ket{j \downn}\leftrightarrow \ket{0 \upp}$ at a rate $2 \Omega \approx 2 \Omega_0 |x_0 x_j|$. We may find $\omega_B^{(j)}$ such that this is the case, by solving the degree $L+1$ polynomial equation
\ban
\left((1 -  \Omega_0^2 g_j(z) g_0(z))(z - \omega_j) + \Omega_0^2( |x_0|^2  g_j(z) - |x_j|^2 g_0(z))\right)\Big|_{z = \omega_B^{(j)}} = 0
\ean
Effectively, we are adjusting the value of $\omega_B^{(j)}$ so that the even order energy shifts induced by $V$ are canceled out. Note that we need a solution for $\omega_B^{(j)}$ that is contained in $(\omega_j-\Delta/4, \omega_j+\Delta/4)$. Using the fact that $\Omega_0<\Delta/4$, as well as $|g_j|,|g_0|\leq\frac{4}{3 \Delta}$  for $z=\omega_B^{(j)} \in [\omega_j-\Delta/4, \omega_j+\Delta/4]$, it is not difficult to show that the left hand side is negative for $\omega_B^{(j)} = \omega_j - \Delta/4$ and positive for $\omega_B^{(j)} = \omega_j + \Delta/4$. Since the left hand side is smooth over this range, by the Intermediate Value Theorem a root exists within  $(\omega_j-\Delta/4, \omega_j+\Delta/4)$. If our computed root is $\delta \omega_B^{(j)}$ off from the exact solution, from \eqref{Heff1} we see the two states retain a splitting $O(\delta \omega_B^{(j)})$. The necessary accuracy $\delta \omega_B^{(j)}$ can thus be incorporated into the error term $\mb{S}_1 \hat \delta \mb{S}_1$, as long as $\delta \omega_B^{(j)} = O(||\mb{S}_1 \hat \delta \mb{S}_1||) = \Omega_0 |x_0 x_j| \cdot O(r)$.

We therefore assume that $\omega_B^{(j)}$ has been chosen so that the diagonal terms in \eqref{Heff1} are equal at $z = \omega_B^{(j)}$, which means
\ban
 H_{eff}  = \omega_B^* \mb{1} + \Omega \left( \ketbradt{j \downn}{0 \upp\!} + \ketbradt{0 \upp}{j \downn\!}  \right)
\ean
where
$$\Omega = \frac{\Omega_0 x_0 x_j }{1 - \Omega_0^2 g_j g_0} = \Omega_0 x_0 x_j \left(1 + O(r^2) \right)$$
$$\omega_B^* = \omega_B^{(j)} + \Omega_0|x_0 x_j| \frac{|x_0/x_j| \Omega_0 g_j }{1 - \Omega_0^2 g_j g_0} = \omega_B^{(j)} + \Omega \cdot O(r)$$
and we assume that $|x_0/x_j|\leq O(1)$. 

$H_{eff}$ has eigenstates $\ket{v_\pm}=\frac{1}{\sqrt{2}}\left(\ket{j \downn} \pm \ket{0 \upp} \right)$ with energy $\omega_B^* \pm \Omega$, so dynamics under $H_{eff}$ for time $\tau_j = \frac{\pi}{2 \Omega} $ would map state $\ket{j \downn} \rightarrow \ket{0 \upp}$. To show that evolution under $\tilde H_j$ achieves the same mapping, we use Theorem \ref{thm1} and Corollary \ref{cor1} to show it has eigenstates and energies near those of $ H_{eff}$. To do this, we must determine an error bound for $|| \Sigma_P(z) - H_{eff}  ||$.

As in \eqref{sigmaz} above, since $H_{eff} = \Sigma_P(\omega_B^{(j)})$ we have
\ba
\label{sigma_diff}
\Sigma_P(z) - H_{eff} & = & P V \left(\tilde G_Q(z) - \tilde G_Q(\omega_B^{(j)}) \right)  V P 
\ea
where
$$ \tilde G_Q(z) = \frac{Q}{z - Q(H_j + V)Q} = \sum_{s} (z - E_s)^{-1} \ketbrad{\phi_s}$$
for $\ket{\phi_s} \in Q$. Since $|\omega_B^{(j)}-\omega_j|<\Delta/4$, the eigenvalues of $Q H_j Q$ are at least $3 \Delta/4$ away from $\omega_j$. As $||V||<\Delta/4$, by Lemma \ref{lem1} we then conclude that $E_s \in (-\infty, \omega_j-\Delta/2]\cup [\omega_j +  \Delta/2, \infty)$ for all $s$. $\tilde G_Q(z)$ is therefore analytic for $z \in [\omega_j - \Delta/4, \omega_j + \Delta/4]$, so we may compute its Taylor series expansion about $\omega_B^{(j)} \in  (\omega_j - \Delta/4, \omega_j + \Delta/4)$:
$$\tilde G_Q(z) = \sum_s \left[(\omega_B^{(j)} - E_s)^{-1} - (z - \omega_B^{(j)}) (z_s - E_s)^{-2}\right] \ketbradt{\phi_s}{\phi_s\,}$$
For some $z_s \in  [\omega_j - \Delta/4, \omega_j + \Delta/4]$, between $\omega_B^{(j)}$ and $z$. Since $|z_s - E_s|\geq \Delta/4$, we conclude that
\ba
\begin{split}
||\tilde G_Q(z) - \tilde G_Q(\omega_B^{(j)}) ||& \leq |z - \omega_B^{(j)}|\left(\frac{4}{\Delta} \right)^2\\
\implies & \quad  \\
||\Sigma_P(z) - H_{eff}|| &\leq |z - \omega_B^{(j)}| \left(\frac{4 ||V||}{\Delta} \right)^2 \\
 &= |z - \omega_B^{(j)}| (4 r)^2
\end{split}
\label{taylorbound}
\ea

As above, the spectrum of $H_{eff}$ is contained in $[c,d]$, where $c = \omega_B^{(j)} - \Omega\left(1 + O(r)\right)$, $d = \omega_B^{(j)} + \Omega\left(1 + O(r)\right)$. In Theorem \ref{thm1}, we consider only values of $z$ in $[c -\gamma, d + \gamma]$ ($\gamma$ is the error in the eigenvalues of $H_{eff}$, compared to $\tilde H_j$). Thus we can determine $\gamma$ self-consistently by solving
$$\gamma =  |z - \omega_B^{(j)}| (4 r)^2$$
for $z = d + \gamma$ and $z = c - \gamma$. To leading order in $r$, this gives
\beq \gamma = \Omega \cdot O(r)
  \label{epsilon} \eeq
(in fact $\gamma = \Omega \cdot O(r^2)$, but the following result holds for \eqref{epsilon} as well). Applying Theorem \ref{thm1}, we have that the two eigenvalues of $H_{eff}$, $E_{\pm} = \omega_B^* \pm \Omega$, are $\gamma$ close to the eigenvalues of $H_j + V$. The relative error in the energy difference $(E_+-E_-)$ is therefore $O(\gamma/\Omega) = O(r)$. 

Now Corollary \ref{cor1} can be used to show that the eigenvectors of $H_{eff}$ are close to the corresponding eigenvectors of $\tilde H_j$. In the notation of that corollary, we can define $P' = \ketbrad{v_+}$, and see that $\nu = 0, \eta = 2 \Omega$. Denoting the analogous eigenvectors and eigenvalues of $\tilde H_j$ by $\ket{\tilde v_\pm}$ and $\tilde E_{\pm}$, using \eqref{epsilon} we see that 
\ban|\!\bra{\tilde v_+}\! v_+ \rangle |^2 &=& \bra{\tilde v_+} P' \ket{\tilde v_+} \\
&>& \left( 1 - \left(\frac{2 ||V||}{\Delta} \right)^2 \right) \left(1 - \left(\frac{2 \gamma + \nu}{\eta - \gamma} \right)^2\right)\\
& = & 1 - O(r^2)
\ean
and likewise for $\bra{\tilde v_-} \! v_-\rangle$. From this and Theorem 1 we conclude that 
\ba 
\begin{split}
\ket{\tilde v_\pm}& = \ket{v_\pm} + O\left(r \right) = \frac{1}{\sqrt{2}}(\ket{j \downn} \pm \ket{0 \upp}) + O(r)\\
\tilde E_+ - \tilde E_- &=  (E_+ - E_-)\left(1 + O(r) \right) = 2 \Omega(1 +O(r))
\end{split}
\label{error}
\ea

For time evolution $\tau_j = \frac{\pi}{E_+-E_-}= \frac{\pi}{2 \Omega}$, \eqref{error} implies the statement of Lemma \ref{lem4}. To complete the proof, we account for the case when $V \neq \mb{S}_1 V \mb{S}_1$ or $\hat \delta \neq 0$ by including the effect of these terms in $||\Sigma_P(z) - H_{eff}||$. As long as \eqref{epsilon} still holds, we conclude that \eqref{error} is still valid. The full Hamiltonian is now $\tilde H_j = H_j + \mb{S}_1 V \mb{S}_1 + \hat \delta_{eff}$, where $\hat \delta_{eff}$ accounts for the terms we previously neglected. Specifically, 
$$ \hat \delta_{eff} = \hat \delta + (V - \mb{S}_1 V \mb{S}_1)$$
We wish to compute the bound $||\Sigma_P(z) - H_{eff}||$, where now $\Sigma_P(z)$ is defined with respect to the perturbation $\mb{S}_1 V \mb{S}_1 + \hat \delta_{eff}$ (see \eqref{newsigmaz} below). As before, we have $H_{eff} = \Sigma_P(\omega_B^{(j)})|_{\hat \delta_{eff} = 0}$, with $\Sigma_P(z)|_{\hat \delta_{eff} = 0}$ defined as in \eqref{sigmaz}. Suppose $||\Sigma_P(z) - \Sigma_P(z)|_{\hat \delta_{eff}=0}|| = \gamma'$. By the triangle inequality,  
$$||\Sigma_P(z) - H_{eff}|| \leq \gamma' + ||\Sigma_P(z)|_{\hat \delta_{eff} = 0} - H_{eff}||$$
We could then repeat the previous analysis to compute $\gamma$, and get $\gamma = \Omega \cdot O(r) + O(\gamma')$. The results of Theorem \ref{thm1} and Corollary \ref{cor1} could then still be applied to get \eqref{error}, as long as $\gamma'$ also satisfies \eqref{epsilon}.  Below we show that this is the case, as long as \eqref{errorbound} is true. 

Including $\hat \delta_{eff}$ in \eqref{series}, we see that  
\ba
\label{newsigmaz}
\begin{split}
\Sigma_P(z)& =  P H_j P + P V P + P \hat \delta_{eff} P \\
& \quad  + P (\mb{S}_1 V \mb{S}_1+ \hat \delta_{eff}) \frac{Q}{z - Q(H_j + \mb{S}_1V\mb{S}_1 + \hat \delta_{eff})Q } (\mb{S}_1V\mb{S}_1 + \hat
 \delta_{eff})P
\end{split}
\ea
In order to bound all terms proportional to $\hat \delta_{eff}$, we use the relation  $(A - B)^{-1} = A^{-1} +  A^{-1} B  (A-B)^{-1}$ to get
\ban
  \frac{Q}{z - Q ( H_j + \mb{S}_1 V \mb{S}_1 + \hat \delta_{eff}) Q} = \tilde G_{Q}(z) + \tilde G_{Q}(z) \hat \delta_{eff} \frac{Q}{z - Q ( H_j + \mb{S}_1 V \mb{S}_1 + \hat \delta_{eff}) Q}
\ean
where $\tilde G_Q(z) = \frac{Q}{z - Q(H_j + S_1 V S_1)Q}$. This allows us to write:
\ba 
\label{errordiff}
\begin{split}
\Sigma_P(z) -  \Sigma_P(z) |_{\hat \delta_{eff}=0}& =   P \hat \delta_{eff} P  +  P \hat \delta_{eff} \frac{Q}{z - Q(H_j + \mb{S}_1V\mb{S}_1 + \hat \delta_{eff})Q } \hat \delta_{eff} P \\
& \quad  + P \hat \delta_{eff} \frac{Q}{z - Q(H_j + \mb{S}_1V\mb{S}_1 + \hat \delta_{eff})Q } \mb{S}_1V\mb{S}_1 P + \mbox{h.c.} \\
& \quad  + P  \mb{S}_1V\mb{S}_1 \tilde G_Q \mb{S}_1\hat \delta_{eff} \frac{Q}{z - Q(H_j + \mb{S}_1V\mb{S}_1 + \hat \delta_{eff})Q } \mb{S}_1V\mb{S}_1 P 
\end{split}
\ea

We now bound this difference. The operator $ Q (z - H_j - \mb{S}_1 V \mb{S}_1) Q$ can be diagonalized in blocks of $\mb{S}_1$ and $\mb{S}_2$. As before, for $z \in [\omega_j -\Delta/4,\omega_j + \Delta/4]$, within both $\mb{S}_1$ and $\mb{S}_2$ this operator has eigenvalues with magnitude at least $\Delta/2$. In the notation of Lemma \ref{lem3}, we may define $A = Q (z - H_j - \mb{S}_1 V \mb{S}_1) Q$, $B =  Q \hat \delta_{eff} Q$, $G_1 = \Delta/2,G_2 = \Delta/2$, so that $(A - B)^{-1} = \frac{Q}{z - Q(H_j + \mb{S}_1V\mb{S}_1 + \hat \delta_{eff})Q}$. Defining $R_i = ||\mb{S}_i Q \hat \delta_{eff} Q \mb{S}_i ||/\Delta\leq ||\mb{S}_i  \hat \delta_{eff} \mb{S}_i ||/\Delta$, $R_\times = ||\mb{S}_1 Q \hat \delta_{eff} Q \mb{S}_2 ||/\Delta\leq ||\mb{S}_1  \hat \delta_{eff} \mb{S}_2 ||/\Delta$, by Lemma \ref{lem3} one may show that
\ban
&|| \mb{S}_1 \frac{Q}{z - Q(H_j + \mb{S}_1V\mb{S}_1 + \hat \delta_{eff})Q} \mb{S}_2 || & = O(R_\times)\\
&||\mb{S}_1 \frac{Q}{z - Q(H_j + \mb{S}_1V\mb{S}_1 + \hat \delta_{eff})Q} \mb{S}_1 || & = O(1) \\
&||\mb{S}_2  \frac{Q}{z - Q(H_j + \mb{S}_1V\mb{S}_1 + \hat \delta_{eff})Q} \mb{S}_2  || & = O(1)
\ean

Writing all terms of \eqref{errordiff} in $\mb{S}_1$, $\mb{S}_2$ blocks, we have for $z \in [\omega_j -\Delta/4,\omega_j + \Delta/4]$
\ban
P \hat \delta_{eff} &=& \Delta \cdot\l( 
\begin{array}{cc}
O(R_1) & O(R_\times)
\end{array} 
\r) \\
\frac{Q}{z - Q(H_j + \mb{S}_1V\mb{S}_1 + \hat \delta_{eff})Q}
 &=& \Delta^{-1} \cdot \l( 
\begin{array}{cc}
O(1) & O(R_\times)\\
O(R_\times) & O(1)
\end{array} 
\r) \\
\mb{S}_1 V \mb{S}_1 P
 &=& \Delta \cdot \l( 
\begin{array}{c}
 O(r)\\
0 
\end{array} 
\r)
\\
P \mb{S}_1V\mb{S}_1 \tilde G_Q \mb{S}_1 \hat \delta_{eff} &=& \Delta \cdot\l( 
\begin{array}{cc}
O(r R_1) & O(r R_x)
\end{array} 
\r) 
\ean
With these components, using \eqref{errordiff} one may calculate $\gamma' = ||\Sigma_P(z) -  \Sigma_P(z) |_{\hat \delta_{eff}=0}|| = O(\Delta (R_1 +  R_\times^2))  $. We need $||\Sigma_P(z) -  \Sigma_P(z) |_{\hat \delta_{eff}=0}|| =  \Omega \cdot O(r)$, so we require
\ban
R_1 &= &  O(r \cdot \Omega /\Delta)\\
R_\times^2 &= & O(r \cdot \Omega /\Delta)\\
R_2 &\leq& 1/2
\ean
where the last inequality comes from the bound on $\hat \delta_{eff}$ necessary to use Lemma \ref{lem3}. One may check that these statements are satisfied by \eqref{errorbound}. 

\begin{flushright}
$\square$
\end{flushright}
 \noindent{\it Proof of Lemma \ref{lem5}:}

As before, we have $H_j = H_S +H_B$, $\tilde H_j =  H + V + \hat \delta_j$.  Define $P_0$ as the eigenspace of $H$ with energies $\omega_0,\omega_1,...\omega_j$. This corresponds to the space $M_{j-1}^d\subseteq \mb{S}_1$ mentioned in the lemma. In the language of Corollary \ref{cor2}, it corresponds to $\lambda_{-k} = \lambda_{k+}=\omega_k$ and $\Delta$ as defined for $H_S$.  The proof comes in two steps. We define the intermediate Hamiltonian $\tilde H_j' = H + \mb{S}_1 V \mb{S}_1$, with an eigenspace $P_1$ corresponding to energies within $(\omega_0-\Delta/8,\omega_0+\Delta/8)\cup (\omega_1-\Delta/8,\omega_1+\Delta/8) \cup ... \cup (\omega_j-\Delta/8,\omega_j+\Delta/8)$. Likewise, $P_2$ is the eigenspace of $\tilde H_j$ of energies within $(\omega_0-\Delta/4,\omega_0+\Delta/4)\cup (\omega_1-\Delta/4,\omega_1+\Delta/4) \cup ... \cup (\omega_j-\Delta/4,\omega_j+\Delta/4)$. The proof follows by showing that (up to an error $O(r^2)$), any state in $ P_0$ is in $P_1$, and any state in $P_1$ is in $P_2$. This will imply that a state in $P_0$ undergoing evolution $U_j$ will remain in $P_0$. For simplicity of notation, for all equations below let $\ket{v_i}$ represent a normalized state in $P_i$.

Since $\tilde H_j'$ and $H_j$ are block diagonal in $\mb{S}_1$ and $\mb{S}_2$, as long as, $P_1\subseteq \mb{S}_1$ it is sufficient to reduce our analysis to $\mb{S}_1$. This holds if $\mb{S}_2(V + \hat \delta) \mb{S}_2$ does not change the energy of $\mb{S}_2$ states by more than $\Delta/2$, as implied by Lemma \ref{lem1} and \eqref{errorbound}. Considering only $\mb{S}_1$, \eqref{errorbound} implies the bound $|| \mb{S}_1 \left(\tilde H_j' - H\right) \mb{S}_1||/\Delta = O(r)$. By Corollary \ref{cor2}, we have that
\ban
\bra{v_0}P_1\ket{v_0} = 1 - j\cdot O(r^2)
\ean
Writing $\ket{v_0} = a \ket{v_1} + b \ket{v_1^\perp}$ where $P_1\ket{v_1^\perp} = 0$, one can easily show that
\ban
P_1 \ket{v_0} = \sqrt{1 - j\cdot O(r^2)}\ket{v_1}
\ean

Equations \eqref{errorbound} also imply that $P_1$ is energetically separate from $Q_1 = \mb{1} - P_1$ by at least $\Delta' = \Delta/4$, so that $||\tilde H_j - \tilde H_j' ||/\Delta' = O(r)$, and as above,
\ban
P_2 \ket{v_1} = \sqrt{1 - j\cdot O(r^2)}\ket{v_2}
\ean
We can combine these statements to get 
\ban
\bra{v_0}P_2 \ket{v_0} &\geq& \bra{v_0}P_1 P_2 P_1 \ket{v_0}\\
& =& (1 - j\cdot O(r^2)) \bra{v_1}P_2 \ket{v_1}\\
& =& (1 - j\cdot O(r^2))^2
\ean
Finally, writing $\ket{v_0} = a \ket{v_2} + b \ket{v_2^\perp}$, the above statement implies
\ban 
P_2 \ket{v_0} = (1 - j\cdot O(r^2))\ket{v_2}
\ean
and by an identical analysis, for any $\ket{v_2}\in P_2$, there exists $\ket{v_0}\in P_0$ such that
$$P_0 \ket{v_2} = (1 - j\cdot O(r^2))\ket{v_0}$$

Notice that the diagonals of $P_0$ are at least as large as those of $P_2 P_0 P_2$. Since $P_2$ is an eigenspace of $\tilde H_j$, it is clear that $ P_2 U_j = U_j P_2$. Using these facts and the above equalities, we compute the bound:
\begin{flalign*}
\bra{v_0} U_j^\dagger P_0 U_j \ket{v_0} & \geq \bra{v_0} U_j^\dagger P_2 P_0 P_2 U_j \ket{v_0}\\
& = \bra{v_0} P_2 U_j^\dagger P_0 U_j P_2 \ket{v_0}\\
& = (1- j \cdot O(r^2))^2 \bra{v_2}  P_0 \ket{v_2} \\
& = (1- j \cdot O(r^2))^4 = 1 - j \cdot  O(r^2) - j^3 \cdot  O(r^6)\\
& = 1 - j \cdot  O(r^2)
\end{flalign*} 
where in the last line we use the fact that $r \propto L^{-5/2}$. Since $Q_0 = \mb{1}-P_0$ we get
$$|| Q_0 U_j \ket{v_0}||^2 \leq j\cdot O(r^2)$$
Writing $U_j \ket{v_0} = P_0 U_j \ket{v_0} + Q_0 U_j \ket{v_0}$, we conclude the proof noting that $P_0 = M_{j-1}^d$ and that the bound $O(r^2)$ is dependent only on the ratio $\Omega_0/\Delta$.

\begin{flushright}
$\square$
\end{flushright}

\noindent {\it Proof of Lemma \ref{lem6}:}\\
Since the operation $E_j$ starts with a bath measurement and since the $M_j$ projector commutes with the bath projectors $\ketbradts{\!\downn}{\downn}$ and $\ketbradts{\!\upp}{\upp}$, we may assume without loss of generality that 
\ban
\rho = M_j^d \rho M_j^d + p_0 \ketbrad{0 \upp}
\ean
where $M_j^d = \mbox{Span}\{\ket{k \downn}\}_{k=0}^j$. Furthermore, since 
\ban
E_j( \ketbrad{0 \upp}) = \ketbrad{0 \upp} = M_{j-1} E_j( \ketbrad{0 \upp}) M_{j-1}
\ean
by the linearity of TCP maps it suffices to analyze the component of $\rho$ within $M_{j}^d$. We may therefore assume that $\rho = M_j^d \rho M_j^d$. Since $\rho$ is a density matrix, we have that 
\ban
\rho = \sum_{l} p_l \ketbrad{v_l \downn}
\ean
where $\ket{v_l \downn} \in M_{j}^d$ and $l$ is a sum over at most $ j+1 = \mbox{dim}(M_{j}^d)$ terms. Each $\ket{v_l\downn}$ may be decomposed into components parallel and orthogonal to $\ket{j \downn}$:
\ban
\ket{v_l \downn} = a_l \ket{j \downn} + b_l \ket{v_l^\perp}
\ean
where $\ket{v_l^\perp}\in M_{j-1}^d$. By Lemma \ref{lem5}, we have that
\ban
U_j \ket{v_l^\perp} = M_{j-1}^d U_j \ket{v_l^\perp} + \sqrt{j}\cdot O(r)
\ean
Likewise by Lemma \ref{lem4}, 
\ban
U_j \ket{j \downn} & = &\ket{0 \upp} + O(r) \\
& = & M_{j -1} U_j \ket{j \downn} + O(r)
\ean
Since $M_{j-1}^d \subset M_{j-1}$, we conclude that
\ban
U_j \ket{v_l \downn} =M_{j-1} U_j \ket{v_l \downn} + \sqrt{j}\cdot O(r)
\ean
Finally, since $E_j$ is a linear operator, we see that
\ban
E_j(\rho) &=& \sum_{l} p_l E_j\left(\ketbrad{v_l \downn}\right)\\
& = &  \sum_{l} p_l \left(U_j \ketbrad{v_l \downn} U_j^\dagger \right)\\
& = &  \sum_{l} p_l \left( M_{j-1} U_j \ket{v_l \downn} + \sqrt{j}\cdot O(r) \right)\\
& \quad & \cdot \left(   \bra{v_l \downn} U_j^\dagger M_{j-1} +\sqrt{j}\cdot  O(r) \right)\\
& =& M_{j -1} E_j(\rho) M_{j-1} + \hat R_j
\ean
where $\hat R_j$ is the sum of all terms proportional to $O(r)$. Using the triangle inequality and the fact that the $p_l$ sum to 1, we see that $||\hat R_j||= \sqrt{j}\cdot O(r)$. Since $\hat R_j$ is a sum of at most $(j+1)$ operators, each of rank 2, we see that $\mbox{rank}(\hat R_j)\leq 2(j+1)$. Therefore $|\mbox{Tr}[\hat R_j]| \leq \mbox{rank}(\hat R_j)\cdot||\hat R_j|| = j^{3/2}\cdot O(r)$. Since a projection of a density matrix is proportional to a density matrix, we may write $M_{j -1} E_j(\rho) M_{j-1} = \lambda_j \rho_{j-1}$, with $\lambda_j = \mbox{Tr}[M_{j-1}E_j(\rho) M_{j-1}] = \mbox{Tr}[E_j(\rho) - \hat R_j] = 1 - j^{3/2}\cdot O(r)$.
\begin{flushright}
$\square$
\end{flushright}

\noindent {\it Proof of Lemma \ref{lem7}:}\\
$\mbb{E}$ is defined by the chain of TCP maps, 
\ban
\mathbb{E} = E_1\circ E_2 ... \circ E_L
\ean
The initial state of the system and bath is described by the density matrix $\rho_L = \ketbrads{F \downn}$, where $\ket{F \downn}\in M_L$. Repeated application of Lemma \ref{lem6} gives
\ban
\mathbb{E}(\rho_L) &=&  E_1\circ E_2 ... \circ E_{L-1}\left(\lambda_L \rho_{L-1} + \hat R_L \right)\\
& = & E_1\circ E_2 ... \circ E_{L-2}\left( \lambda_L \lambda_{L-1} \rho_{L-2} + \lambda_L \hat R_{L-1} + E_{L-1}(\hat R_L )\right)\\
& \vdots &\\
& = & \left(\prod_{k = 1}^L  \lambda_k\right) \rho_0 + \hat R_{tot}
\ean
where $M_0 \rho_0 M_0 = \rho_0$, $\hat R_{tot}$ represents all other terms, and $\mbox{Tr}[\hat R_{tot}] = 1 - \left(\prod_{k = 1}^L  \lambda_k\right)$ since $\mathbb{E}$ is trace-preserving.

We will bound the infidelity, $ 1 - \mbox{Tr}[M_0 \mathbb{E}(\rho_L)] =  1 - \left(\prod_{k = 1}^L  \lambda_k\right) - \mbox{Tr}[M_0 \hat R_{tot}]$, by showing that $1 - \left(\prod_{k = 1}^L  \lambda_k\right)$ and $\mbox{Tr}[M_0 \hat R_{tot}]$ are small. First, consider the quantity $y =\log\left(\prod_{k = 1}^L  \lambda_k\right) = \sum_{k = 1}^L \log(\lambda_k)$. By Lemma \ref{lem6}, we see that $\lambda_k = 1 -\mbox{Tr}(\hat R_k)$. Given that $|\log(1 - x)| \leq 2 |x|$ for $|x|<1/2$, we conclude that for $|\mbox{Tr}(\hat R_k)|<1/2$,
\ban
|y| &\leq& \sum_{k = 1}^L \left|\log(1 - \mbox{Tr}[\hat R_k])\right|\\
& \leq & \sum_{k = 1}^L 2 |\mbox{Tr}[\hat R_k]| \\
& = & \sum_{k = 1}^L k^{3/2} \cdot O(r)\\
& = & L^{5/2}\cdot O( r)
\ean
Thus, to leading order in $r$, 
\ban
 1 - \left(\prod_{k = 1}^L  \lambda_k\right) = 1- e^y =  L^{5/2} \cdot O( r)\\
\ean

To show the second term is small, we must bound $\hat R_{tot}$, which is the sum of all error terms:
\ban
\hat R_{tot} = \sum_{k=1}^{L} \left(\prod_{j = k+1}^L \lambda_j \right) E_1 \circ ...\circ E_{k-1} ( \hat R_k )  
\ean
From the simple form of $E_j$ (see \eqref{Ej}), we see that 
\ban
E_1 \circ ...\circ E_{k-1} ( \hat R_k )  =  \sum_{j=1}^{k} A_j \hat R_k A_j^\dagger
\ean
where 
\ban
A_j = \ketbrads{\uparrow} \left(U_{j} \ketbrads{\downarrow} \right)\cdot \left(U_{j+1} \ketbrads{\downarrow} \right)\cdot ... \cdot  \left(U_{k-1} \ketbrads{\downarrow} \right)
\ean
for $2\leq j \leq k$, and
\ban
A_1 &=& \left(U_{1} \ketbrads{\downarrow} \right)\cdot \left(U_{2} \ketbrads{\downarrow} \right)\cdot ... \cdot  \left(U_{k-1} \ketbrads{\downarrow} \right) \\
A_k &=& \ketbrads{\uparrow}
\ean
Since $A_j$ is a product of projectors and unitaries, we must have that $||A_j \hat R_k A_j^\dagger|| \leq ||\hat R_k||$, so by the triangle inequality and Lemma \ref{lem6} it follows that
 $$
||E_1 \circ ...\circ E_{k-1} ( \hat R_k )|| \leq k ||\hat R_k ||=k ^{3/2} \cdot  O( r) \implies$$ 
$$|| \hat R_{tot}||  = \sum_{k=1}^L k^{3/2} \cdot  O(r) = L^{5/2} \cdot O(r)$$

Finally, we note that since $M_0$ is a projector of rank two, $ M_0 \hat R_{tot} M_0$ also has rank at most 2. By the cyclic property of the trace, we conclude that
$$ |\mbox{Tr}[M_0\hat R_{tot} ]| = |\mbox{Tr}[M_0\hat R_{tot}M_0 ] |\leq \mbox{rank}(M_0\hat R_{tot} M_0) \cdot ||M_0 \hat R_{tot} M_0 || = L^{5/2} \cdot O(r)$$
Combining the two results, we have that
\ban
 1 - \mbox{Tr}[M_0 \mathbb{E}(\rho_L)]  =L^{5/2}\cdot  O( r)
\ean
Thus, as long as $r = O(\epsilon/ L^{5/2})$, the algorithm succeeds with infidelity $O(\epsilon)$.
\begin{flushright}
$\square$
\end{flushright}

\section{Extension Analysis}

We now discuss an augmentation of the previous cooling technique which does not require knowledge of the overlaps $x_k$ describing the fiducial state, $\ket{F \downn}$. It is described in detail in the article, though we summarize it here. We assume that the system Hamiltonian $H_S$ has the form
$$ H_S = P_1 H_S  P_1+ P_2 H_S P_2$$
where $P_1$ is the eigenspace of $H_S$ with energy between $(\omega_1-\delta \omega)$ and $(\omega_1 + \delta \omega)$, $\ket{0}$ is the nondegenerate groundstates $H_S$ with energy $\omega_0 =0$, and $P_2$ is a projector into the space orthogonal to $\mbox{Span}\{\ket{0}\}\oplus P_1$. To relate to notation in the previous section, we define the projectors $\mb{S}_1 = (\ketbrad{0}+P_1)\otimes \mb{1}$, ${S}_2 = P_2 \otimes \mb{1}$. Since these act trivially on the bath, as a slight abuse of notation we will sometimes refer to $\mb{S}_1$ and $\mb{S}_2$ as operating on the system Hilbert space alone. We define the spectral gap between $\ket{0},P_1$ and $P_2$:
\ba
\label{3stategap}
\Delta = \min\l\{ \omega_1, E, |E-\omega_1| : E \in \mbox{Spec}(H_S|_{\mb{S}_2}) \r\}
\ea

In full, the unperturbed Hamiltonian is
\ba
\label{H3state}
\begin{split}
H & = H_S \otimes \left(\ketbrad{C} + \ketbrad{R} -\ketbrad{L} \right)\\
& \quad + \omega_1 \mb{1}_S \otimes \left( \ketbrad{R} + \ketbrad{L} \right)
\end{split}
\ea
where $\ket{C}, \ket{R}$ and $\ket{L}$ are orthogonal basis vectors for the bath Hilbert space. We start by preparing a fiducial state,
\ban
\ket{F\,  C} = f_1 \ket{1\, C} + f_\perp \ket{F_\perp \, C}
\ean
where $\ket{1}\in P_1$, $\bra{1} F_\perp \rangle = 0$, and we are given a lower bound for $|f_1|$. The algorithm proceeds by simulating the evolution of Hamiltonians $H + X$, where $X$ satisfies
\ba
\begin{split}
\label{V3}
X  &= T_S \otimes \left( \ketbradt{C}{B} + \ketbradt{B}{C} \right)\\
 \ket{B} & =  \frac{1}{\sqrt{2}} ( \ket{L} + \ket{R})\\
\Omega \ket{1} & = P_1 T_S \ket{0}
\end{split}
\ea
where by phase convenstion $\Omega$ is real. Again, although $||T_S|| = \Omega_0$ is a known quantity, we are only given a lower bound $\Omega^*$ for $\Omega$. 

The algorithm is probabilistic, and involves a single evolution step for time $\tau \sim \frac{1}{\Omega}$, followed by a measurement of the bath. If the bath is measured in state $\ket{B}$, then the desired transition $\ket{1 \,C}\rightarrow \ket{0 \, B}$ could have occured. We verify this by applying the bath unitary $\ket{B}\leftrightarrow\ket{L},\ket{D}\leftrightarrow\ket{R}$, then evolving under $H + Y$ for time $\tau = \frac{\pi}{2 \Omega_0}$ where
\ba
Y = \Omega_0 \mb{1}_S \otimes \left(\ketbradt{L}{R} + \ketbradt{R}{L} \right)
\ea
A measurement of a bath transition $\ket{L}\rightarrow \ket{R}$ would indicate that the system is in its ground state, while in all other cases a transition $\ket{\psi L}\rightarrow \ket{\psi R}$ is suppressed by energy conservation. If either the first or second bath measurements fail, we reinitialize the system and start again. 

As before, we again show that given some bounds, the algorithm is robust against simulation errors and coupling between $\mb{S}_1$ and $\mb{S}_2$:
\ba
\begin{split}
\label{3errorbound}
& r = \Omega_0/\Delta < 1/8\\
& \delta \omega < r \cdot \Omega(V) \\
&\frac{|\bra{0} T_S \ket{0}|^2}{\omega_1} < r \cdot \Omega(V) \\
  &\frac{||\mb{S}_1 \hat \delta \mb{S}_1||}{\Delta} <  r \cdot\frac{\Omega(V)}{\Delta}\\
&\frac{|| \mb{S}_1(V + \hat \delta) \mb{S}_2 ||^2 }{\Delta^2} <  r \cdot\frac{\Omega(V)}{\Delta}\\
&\frac{||\mb{S}_2(V + \hat \delta) \mb{S}_2||}{\Delta} < \Delta/2 
\end{split}
\ea
where $\Omega(V) = \Omega_0$ for $V = Y$, and $\Omega(V) = \bra{1}T_S\ket{0} = \Omega $ for $V = X$. As seen below, for the probabilistic scheme to succeed with fidelity $1 - O(\epsilon)$, we must scale $r$ as $ O(|f_1| \epsilon^{3/2})$.

\begin{lemma}[Fidelity of the Unitary Evolutions]
\label{lem8}
Let $U(\tau) = e^{- i \tau (H + X + \hat \delta)} $ and assume \eqref{3errorbound}. Then
\ba
\label{Uteq}
U(\tau)\ket{1 \, C} = \cos(\phi_t) \ket{1\, C} - i \sin(\phi_t) \ket{0\, B} + O(r)
\ea
where $\phi_t = \tau \Omega(1 + O(r)) $.  The error term in $\phi_t$ and in \eqref{Uteq} is uniform over $\tau$. Likewise, let $U_v(\tau) = e^{- i \tau (H + Y + \hat \delta)} $. Then
\ba
U_v(\tau) \ket{0\, L} = \cos(\phi_v) \ket{0\, L} - i \sin(\phi_v) \ket{0\, R} + O(r)
\ea
where $\phi_v = \tau \Omega_0 $.
\end{lemma}
\\
\begin{lemma}[Verification Step]
\label{lem9} 
\ba
\begin{split}
\max_{\ket{\psi}}\bra{\psi \, L} U_v^\dagger \left(\mb{1}_S\otimes \ketbrad{R}\right) U_v \ket{\psi\, L} &= &O(r^2)\\
\max_{\ket{\psi}}\bra{\psi \, R} U_v^\dagger \left(\mb{1}_S\otimes \ketbrad{L}\right) U_v \ket{\psi\, R} &= &O(r^2)
\end{split}
\ea
where the maximum is taken over all normalized system states $\ket{\psi}$ such that $\bra{\psi} 0 \rangle = 0$.
\end{lemma}
\\
\begin{theorem}[Success of the Probabilistic Scheme]
\label{thm4}
 Say that $r = O(|f_1| \epsilon^{3/2})$ as $\epsilon \rightarrow 0^+$, and that the $U_t$ simulation time $\tau $ is sampled randomly within the range $[\frac{\pi}{\Omega^*}, \frac{2 \pi}{\Omega^*} ]$, where $\Omega^*< \Omega$. Then the verification step accepts with probability $p_{v} = \frac{1}{|f_1|^2}\cdot O(1)$. Given an acceptance, the probability of the system being in its ground state is $p_{success} = 1 - O(\epsilon)$. Since $\Omega < r \Delta$, the average simulation time $\mean{T}$ satisfies
\ban
o\left(\frac{1}{|f_1|^3} \frac{1}{ \epsilon^{3/2} \Delta}\right) = \langle T \rangle = O\left(\frac{1}{|f_1|^2} \frac{1}{ \Omega^*}\right)
\ean
\end{theorem}

The proof of Lemma \ref{lem8} is analogous to the proof of Lemma \ref{lem4}. We first analyze the success of the unitary evolutions under the assumption the most unwanted terms are zero, and show that it leads to the desired outcome. We then bound the effect of the unwanted terms on the unitary evolution.\\ 
\noindent {\it Proof of Lemma \ref{lem8}:}

We begin by proving the first statement of the lemma, for $V = X = T_S \otimes \l(\ketbradt{B}{C} + \ketbradt{C}{B} \r)$. As in the proof of Lemma \ref{lem4}, instead of analyzing $H + V + \hat \delta$ we start by looking at the evolution of $H + \mb{S}_1 V \mb{S}_1 $, then obtain a bound on the errors caused by $\hat \delta_{eff} = V - \mb{S}_1 V \mb{S}_1 + \hat \delta$. Define $P$ as the eigenspace of $H$ with energy in $[\omega_1-\Delta/4,\omega_1+\Delta/4]$. Notice that the projector $P$ is $P = P_1\otimes \ketbrad{C} + \ketbrad{0} \otimes( \ketbrad{D} + \ketbrad{B})$, where $\ket{B} = \frac{1}{\sqrt{2}}(\ket{L} + \ket{R})$, $\ket{D} = \frac{1}{\sqrt{2}}(\ket{L} - \ket{R})$, and that $P\subset \mb{S}_1$. Before we calculate the self energy operator $\Sigma(z)$ at $z = \omega_1$, we note the following relations:
\ban
P & = & \ketbrad{0}\otimes(\ketbrad{ L} + \ketbrad{ R}) + P_1 \otimes \ketbrad{ C}\\
& = & \ketbrad{0}\otimes(\ketbrad{ D} + \ketbrad{ B}) + P_1 \otimes \ketbrad{ C}\\
Q  &=& P_{1}\otimes(\ketbrad{ L} + \ketbrad{ R}) + \ketbrad{0} \otimes \ketbrad{ C} + \mb{S}_2\\
PHP & =&  \omega_1 \ketbrad{0}\otimes(\ketbrad{ D} + \ketbrad{ B}) + P_1 H_S P_1 \otimes \ketbrad{C}\\
QHQ & =&  (P_1 H_S P_1 + \mb{S}_2 H_S \mb{S}_2) \otimes (-\ketbrad{ L} + \ketbrad{ R}) \\
&& + \omega_1 (P_1+ \mb{S}_2)\otimes (\ketbrad{ L} + \ketbrad{ R}) + \mb{S}_2 H_S \mb{S}_2\otimes \ketbrad{C}\\
P V P &=& P \l[T_S \otimes (\ketbradt{C}{B} + \ketbradt{B}{C})\r] P \\
& = & \Omega \l(\ketbradt{1 \, C}{0\, B}+ \ketbradt{0\, B}{1 \, C} \r)
\ean
The next term required in \eqref{series} is the unperturbed Green's function, $G_Q(z) = \frac{Q}{z Q - Q H Q}$. Since $P\subset \mb{S}_1$ and $\mb{S}_1 V \mb{S}_1$, $H$ are both block diagonal in $\mb{S}_1$ and $\mb{S}_2$, we may ignore the $\mb{S}_2$ component of $G_Q(z)$: 
\ban
\mb{S}_1 G_Q(z) \mb{S}_1 &=& \frac{P_1\otimes (\ketbrad{ L} + \ketbrad{ R})}{(z-\omega_1) - P_1H_S P_1\otimes (\ketbrad{R}-\ketbrad{L}) }\\
& \quad &+ \frac{1}{z}\ketbrad{0} \otimes \ketbrad{C}
\ean 
so that
\ban
 \mb{S}_1 G_Q(\omega_1)\mb{S}_1 &=& \frac{P_1}{P_1H_S P_1   }\otimes (\ketbrad{L}-\ketbrad{R})\\
& \quad &+ \frac{1}{\omega_1}\ketbrad{0} \otimes \ketbrad{C}
\ean  

Notice that $\bra{B}\mb{S}_1 G_Q(\omega_1) \mb{S}_1\ket{B} = 0$. From the definition of $\mb{S}_1V \mb{S}_1=  \mb{S}_1 T_S\mb{S}_1 \otimes (\ketbradt{C}{B} + \ketbradt{B}{C})$, we immediately observe that 
\ban
\mb{S}_1 V  \mb{S}_1 G_Q(\omega_1) \mb{S}_1 V\mb{S}_1 = \frac{1}{\omega_1} \mb{S}_1 T_S  \ketbrad{0} T_S \mb{S}_1 \otimes \ketbrad{B}
\ean
Multiplication by $ G_Q(\omega_1) $ again produces a term proportional to $\ket{D}$. Since $V \ket{D} = 0$, this implies that the series \eqref{series} with perturbation $\mb{S}_1 V \mb{S}_1 $ truncates at second order in $V$. Therefore $H_{eff} \equiv \Sigma(\omega_1)$ may be computed to all orders as
\ban
H_{eff} &=& P H P + P V P+ P V \mb{S}_1 G_Q(\omega_1)  \mb{S}_1 V P \\
& = & \omega_1 \ketbrad{0}\otimes(\ketbrad{ D} + \ketbrad{ B}) + P_1 H_S P_1 \otimes \ketbrad{C} \\
&& + \Omega \l(\ketbradt{1 \, C}{0\, B}+ \ketbradt{0\, B}{1 \, C} \r) \\
& &+  \frac{|\!\bra{0}T_S \ket{0}\!|^2}{\omega_1} \ketbrad{0 B}
\ean

The system Hamiltonian is written $H_S = P_1 H_S P_1 + \mb{S}_2 H_S \mb{S}_2$, where the spectrum of $H_S|_{P_1}$ is contained in $(\omega_1-\delta \omega,\omega_1 + \delta \omega)$. The state $\ket{1} \propto P_1 T_S \ket{0}$ is not necessarily an eigenstate of $H_S$, but $\omega_1^* = \bra{1}H_S \ket{1}$ is contained in $(\omega_1-\delta \omega,\omega_1 + \delta \omega)$. We write the projector into the remainder of $P_1$ as $P_1' = P_1 - \ketbrad{1}$. To see that $H_{eff}$ produces the desired evolution, we rewrite it as 
\ba
\label{Heff3state}
\begin{split}
H_{eff} &= \omega_1 \l( \ketbrad{0\, B}+ \ketbrad{1 \, C} \r) + \l(\omega_1 \ketbrad{0\, D} + P_1' H_S P_1' \otimes \ketbrad{C}\r)\\
& +\Omega \l(\ketbradt{1 \, C}{0\, B}+ \ketbradt{0\, B}{1 \, C} \r) \\
& + \l( (\omega_1^*-\omega_1)\ketbrad{1}+P_1' H_S \ketbrad{1} + \mbox{h.c.}\r) \otimes \ketbrad{C} + \frac{|\!\bra{0}T_S \ket{0}\!|^2}{\omega_1} \ketbrad{0 B}
\end{split}
\ea

Observe that if we neglect the terms on the third line of \eqref{Heff3state}, $H_{eff}$ has eigenvectors $\ket{v_\pm} = \frac{1}{\sqrt{2}}\l(\ket{1 C} \pm \ket{0 B} \r)$ with eigenvalues $\omega_1 \pm \Omega$, which exactly produce the desired evolution \eqref{Uteq}. Since $||P_1'(H_S - \omega_1 P_1') P_1'|| \leq ||P_1(H_S - \omega_1 P_1) P_1|| \leq \delta \omega$, by Lemma \ref{lem1} all other eigenvalues of the approximate $H_{eff}$ are in $(\omega_1 - \delta \omega, \omega_1 + \delta \omega)$. The eigenvalues $\omega_1 \pm \Omega$ are therefore non-degenerate and energetically separated from the rest of the spectrum by a gap $\Omega-\delta \omega$. This fact will allow us to use Theorem \ref{thm2} below to show that, up to an error of order $O(r)$, $\ket{v_\pm}$ correspond to eigenvectors of $H_{eff}$.

The terms in the third line of \eqref{Heff3state} are bounded by $\Omega \cdot O(r)$. To see this, note that $\frac{|\!\bra{0}T_S \ket{0}\!|^2}{\omega_1} \leq \Omega\cdot O(r)$ is already an explicit assumption. The bound for $\l( (\omega_1^*-\omega_1)\ketbrad{1}+P_1' H_S \ketbrad{1} + \mbox{h.c.}\r) = P_1( H_s - \omega_1 P_1) P_1 - P_1'( H_s - \omega_1 P_1') P_1'$ comes from the fact that $||P_1'(H_S - \omega_1 P_1') P_1'|| \leq ||P_1(H_S - \omega_1 P_1) P_1|| \leq \delta \omega = \Omega \cdot O(r)$. By invoking Lemma \ref{lem1} and Theorem \ref{thm2} we conclude that $H_{eff}$ has eigenvectors $\ket{\tilde \pm}$ with eigenvalues $\omega_1 \pm \Omega\cdot (1 + O(r))$ such that $|\bra{\pm} \tilde \pm \rangle|^2 \geq 1 - O(r^2)$, and that the rest of the spectrum of $H_{eff}$ is $\Omega\cdot (1 + O(r))$ away from these eigenvalues. 

The rest of the proof of \eqref{Uteq} is now identical to the argument in Lemma \ref{lem4}. Using the bound for $\Omega_0 = ||T_S||$, in the case when $\hat \delta_{eff} = 0$ we bound $||\Sigma(z)|_{\hat \delta_{eff}} - H_{eff}||$ for $z \in [\omega_1-\Omega\cdot(1 + O(r)),\omega_1+ \Omega\cdot(1 + O(r))]$ using the Taylor's expansion of $\frac{Q}{z Q - Q(H+\mb{S}_1 V \mb{S}_1)Q}$, . Then, using Lemma \ref{lem3} we bound the error in $\Sigma(z)$ obtained by neglecting $\hat \delta_{eff} = V - \mb{S}_1 V \mb{S}_1 + \hat \delta$, and show that it is equal to $\Omega \cdot O(r)$ under our assumed bounds \eqref{3errorbound}. Since $||\Sigma(z) - H_{eff}||$ is still sufficiently small, we conclude by Theorem \ref{thm1} and corollary \ref{cor1} that $H+V+\hat \delta$ has eigenvalues $\omega_1 \pm \Omega\cdot (1 + O(r))$, and that these eigenvalues correspond to $\frac{1}{\sqrt{2}}\l(\ket{1 \,C} + \ket{0\, B} \r) + O(r)$.

The proof of the second statement is nearly identical to the first. Noting that now $V = \Omega_0 \mb{1}_S \otimes (\ketbradt{R}{L} + \ketbradt{L}{R})$, we have $PVP = \Omega_0 \ketbradt{0 R}{0 L} + \mbox{h.c.}$ and $PVQ = 0$. Assuming that $\hat \delta = 0$, the level shift operator $\Sigma(z)$ is now exactly equal to $PHP + PVP$, so $H_{eff} = \Sigma(\omega_1)$ satisfies
\ban
H_{eff} &= &\omega_1 \l( \ketbrad{0\, L}+ \ketbrad{0 \, R} \r) + P_1 H_S P_1 \otimes \ketbrad{C}\\
& +&\Omega_0 \l(\ketbradt{0 \, L}{0\, R}+ \ketbradt{0\, R}{0 \, L} \r)
\ean
which clearly has eigenvalues $\omega_1 \pm \Omega_0$ corresponding to $\frac{1}{\sqrt{2}}(\ket{0 L} \pm \ket{0 R})$, with the rest of its spectrum in $(\omega_1-\delta \omega,\omega_1 + \delta \omega)$. $H_{eff}$ therefore produces the desired evolution. The rest of the proof, in which we bound $||\Sigma(z)-H_{eff}||$, again continues in the same way as in Lemma \ref{lem4}, with the substitution of $\Omega_0$ in place of $\Omega$.

\begin{flushright}
$\square$
\end{flushright}

\noindent {\it Proof of Lemma \ref{lem9}:}\\
The proof of this analogous to the proof of Lemma \ref{lem5}. As before, let $P$ represent the eigenspace of $H$ with energy contained in $(-\infty,\omega_1-\Delta]$. Notice that $P$ corresponds only to bath states in state $\ket{C}$ or $\ket{L}$. $Q = \mb{1}-P$ corresponds to the eigenspace of energies within $[\omega_1,\infty)$. In the language of Theorem \ref{thm2}, we have $\lambda_- = -\infty, \lambda_+ = \omega_1-\Delta/2$, and $\Delta$ defined as in \eqref{3stategap}.

Given \eqref{3errorbound}, by the triangle inequality we conclude that $||Y + \hat \delta||/\Delta = O(r)$. Define $\tilde P$ as the eigenspace of $\tilde H_ = H + Y + \hat \delta$ with energy in $(\lambda_-,\lambda_+) = (-\infty,\omega_1-\Delta/2)$. By Theorem \ref{thm2}, for any state $\ket{ v} \in P$, we see that 
\ban
\bra{ v}\tilde P \ket{ v}& >&  1 - \left(\frac{2||\tilde H - H||}{\Delta} \right)^2\\
& = & 1 - O(r^2)
\ean
This implies that $\tilde P \ket{ v} = \sqrt{1-O(r^2)} \ket{\tilde v}$, where $\ket{\tilde v}$ represents an arbitrary normalized vector in $ \tilde P$. Likewise, for $\ket{\tilde  v}\in \tilde  P$, $\bra{\tilde v} P \ket{\tilde v} = 1 - O(r^2)$. Since $\tilde P$ is an eigenspace of $\tilde H$, for $U_v = \exp(-i \tau \tilde H )$, $\tilde P U_v = U_v \tilde P$. Noting that $\bra{\psi}\tilde P P \tilde P \ket{\psi} \leq \bra{\psi} P \ket{\psi}$ for all $\ket{\psi}$, we conclude that for any $\ket{v} \in P$, 
\ban
\bra{v} U_v^\dagger P U_v \ket{v} &\geq& \bra{v} U_v^\dagger \tilde P P  \tilde P U_v \ket{v} \\ 
& = & \left(\bra{v}\tilde P U_v^\dagger\right) P \left(U_v \tilde P \ket{v}\right)\\
& = & (1 - O(r^2))\bra{\tilde v} P \ket{\tilde v}\\
& = & (1 - O(r^2))^2 = 1 - O(r^2)
\ean

Let $\ket{\psi\, L}$ be given such that $\bra{\psi} \! 0\rangle = 0$. By examinig the spectrum of $H$, one sees that $\ket{\psi\, L} \in P$, and that the eigenspace $\mathbb{H}_S\otimes \mbox{Span}\{\ket{R} \}$ is contained within $Q$, so that the operator $\mb{1}_S\otimes \ketbrad{R} \leq Q = (\mb{1} - P)$. From the above inequality we conclude that
\ban
\bra{\psi \, L}U_v^\dagger \left( \mb{1}_S\otimes \ketbrad{R} \right) U_v \ket{\psi \, L} &\leq & \bra{\psi \, L}U_v^\dagger \left(\mb{1} - P \right) U_v \ket{\psi \, L} \\ 
& = & 1 -  \bra{\psi \, L}U_v^\dagger  P  U_v \ket{\psi \, L}\\
& = & O(r^2)
\ean
By an identical argument, we may prove the second statement of the claim as well.
\begin{flushright}
$\square$
\end{flushright}

\noindent {\it Proof of Theorem \ref{thm4}:}\\
Given initial state $\ket{F C}$ and time evolution $U_t(\tau)$ (as defined in Lemma \ref{lem8}), the probability of a verification event is given by
\ban
p_v = \bra{F C} U_t^\dagger X^\dagger X U_t \ket{F C}
\ean
where 
\ban
X = \ketbrad{R} U_v \ketbradt{L}{B}
\ean
and $U_v$ is evaluated for time $\frac{\pi}{2 \Omega_0}$. Likewise, the probability of the system being in the ground state after the verification has occurred is
\ban
p_{vs} = \bra{F C} U_t^\dagger X^\dagger \ketbrad{0}X U_t \ket{F C}
\ean
where $\ket{0}$ is the ground state of $H_S$. Along with finding $p_v$, we wish to calculate the success probability of the algorithm conditional on a verification event, $p_{s|v} = p_{v s}/p_v$. These three probabilities are functions of the parameter $\phi$, where $\phi/\tau = \Omega(1 + O(r))$ is half the energy splitting of the eigenstates $\ket{\tilde v_{\pm}} \approx \frac{1}{\sqrt{2}}(\ket{1 C} \pm \ket{0 B})$ used in the evolution $U_t(\tau)$. 

By the first result of Lemma \ref{lem8}, for any time evolution $U_t(\tau)$ we may write
\ban
U_t(\tau) \ket{F \, C} = f_1 \left(\cos(\phi_t) \ket{1\,C} - i \sin(\phi_t) \ket{0 \, B} + O(r) \right) + f_\perp U_t \ket{f_\perp C}
\ean
Since Lemma \ref{lem8} implies coherent oscillations between $\ket{1\, C}$ and $\ket{0 \, B}$, it must be that $\bra{0 \, B} U_t(\tau) \ket{f_{\perp} C} = O(r)$, so that $\bra{0 B} U_t(\tau) \ket{F C} = -i (f_1 \sin(\phi_t) + O(r))$. We mention that the bound $O(r)$ is independent of $\tau$, i.e. as $r\rightarrow 0^+$ there is a constant $c>0$ such that $|\bra{0 B} U_t(\tau) \ket{F C} + i (f_1 \sin(\phi_t)|< c \cdot r$ for all $\tau$.

Likewise, the second result of Lemma \ref{lem8} implies that
\ban
U_v\l(\tau =  \frac{\pi}{2 \Omega_0}\r) \ket{0\, L} = \ket{0 \,R} + O(r)
\ean
so
\ban
X \ket{0\, B} = \ket{0 \, R} + O(r)
\ean
For system states $\ket{\psi}$ such that $\bra{\psi} 0 \rangle = 0$, by Lemma \ref{lem9} we have that 
\ban
\max_{\langle \psi \ket{\psi} = 1 }\bra{\psi\, L} U_v^\dagger \ketbrad{R} U_v \ket{\psi \,L} = O(r^2)
\ean
and since $U_v^\dagger \ketbrad{R} U_v $ is a projector, it must be that $ U_v^\dagger \ketbrad{R} U_v \ket{\psi \,L} = O(r)$, where the error bound is uniform over all $\ket{\psi}$. We conclude that for all composite states $\ket{\psi_{SB}}$ such that $\bra{ \psi_{SB}} 0 B \rangle = 0$,  
\ban
X \ket{\psi_{SB}} = U_v \l(U_v^\dagger \ketbrad{R} U_v\r) \cdot \l(\ketbradt{L}{B} \psi_{SB} \rangle \r)=  O(r)
\ean
and that this bound is the same over all such $\ket{\psi_{SB}}$.
Since $\bra{0 B} U_t(\tau) \ket{F C} = -i (f_1 \sin(\phi_t) + O(r))$, we may write
\ban
 U_t(\tau) \ket{F \, C} = -i (f_1 \sin(\phi_t) + O(r)) \ket{0 \, B} + c(\tau) \ket{\psi_{SB}(\tau)}
\ean
where $|c(\tau)|<1$ and $\bra{ \psi_{SB}(\tau)} 0 B \rangle = 0$ for all $\tau$.

Applying $X$ gives
\ban
X U_t(\tau) \ket{F \, C} &=& -i (f_1 \sin(\phi_t) + O(r)) \ket{0 \, R} + O(r) \\
& = &  -i (f_1 \sin(\phi_t) + A(\phi_t)) \ket{0 \, R} + B(\phi_t) \ket{\psi_\perp(\tau)\, R}
\ean
where $\bra{\psi_\perp(\tau)} 0 \rangle = 0$ for all $\tau$. As the bounds derived from Lemma \ref{lem8} and 9 are independent of $\tau$, we have that $A = \max |A(\phi_t)| = O(r)$ and $B = \max |B(\phi_t)| = O(r)$. We may now directly compute $p_v$ and $p_{vs}$:
\ban
p_{v}(\phi_t)& =& |f_1 \sin(\phi_t) + A(\phi_t)|^2 + |B(\phi_t)|^2\\
p_{vs }(\phi_t) & = & |f_1 \sin(\phi_t) + A(\phi_t)|^2 
\ean
so the success probability for a given $\phi_t$ is
\ban
p_{s|v}(\phi_t) = \frac{|f_1 \sin(\phi_t) + A(\phi_t)|^2}{|f_1 \sin(\phi_t) + A(\phi_t)|^2+ |B(\phi_t)|^2}
\ean

Suppose that when sampling over values of $\phi_t$, with probability at least $1 - \epsilon/2$ we have $ p_{s|v}(\phi_t)>(1-\epsilon/2)$. Then $p_{success} > (1-\epsilon/2)^2 > 1 - \epsilon$ for $0<\epsilon<1$, which is the desired result. In terms of the relation above, this condition equivalent to
\ban
|f_1 \sin(\phi_t) + A(\phi_t)| > |B(\phi_t)|\sqrt{\frac{2}{\epsilon} -1}
\ean
As $|f_1 \sin(\phi_t) + A(\phi_t)|> |f_1 \sin(\phi_t)| - A$ and $|B(\phi_t)|<B$, this relation is satisfied if
\ba
\label{bound}
|\sin(\phi_t)| \geq \frac{A + B\sqrt{\frac{2}{\epsilon}}}{|f_1|} = O\l(\frac{r}{\sqrt{\epsilon} |f_1|}\r)
\ea

Hence we obtain an infidelity at most $\epsilon$ as long \eqref{bound} is violated with probability at most $\epsilon/2$. Note that if $\phi_t$ is sampled uniformly over a range larger than $\pi/2$ (as ensured by our sampling scheme for $\tau$), the probability that $|\sin(\phi_t)|<c$ for some small number $c>0$ is $p_{fail} = O(c)$ as $c \rightarrow 0$. We require $p_{fail}<\epsilon/2$ in \eqref{bound}, which is satisfied for $r = O(|f_1| \epsilon^{3/2})$. This gives success the bound $p_{success} > 1 - \epsilon$ stated in the Lemma. Using the same argument, we see that to have $|f_1 \sin(\phi_t)|>2 A = O(r)$ with probability at least $1/2$, we only require $r = O\l(|f_1| \r)$, so under the more stringent scaling we may also conclude that the verification probability $p_v(\phi_t)$ is greater than $|f_1|^2/4$ with probability $O(1)$. This gives the desired scaling, $p_{accept} = O\l(\frac{1}{|f_1|^2}\r)$.

\begin{flushright}
$\square$
\end{flushright}

\section{Universality}
The following is derived from results in \cite{Kitaev2002,Kempe2004,Aharonov2007}. Using the notation of \cite{Kempe2004}, $\ket{0^n}$ represents a state of $n$ qubits, each initialized in the qubit state $\ket{0}$. The universality of both QSC schemes follows immediately from this claim, and the fact that 1 and 2-local unitaries are universal \cite{Barenco1995}.\\

\begin{theorem}[Universality of QSC]
\label{thm5}
Let $\mb{U} = U_L U_{L-1} ... U_{1}$ be composed of $L$ one and two-qubit gates on $n$ qubits. There exists a 5-local Hamiltonian $H_S$ on $n + L$ qubits, whose ground state $\ket{\eta}$ tracks the history of the unitary evolution:
$$\ket{\eta} = \sum_{l=0}^L U_{l}U_{l-1}...U_1 \ket{0^n}\ket{1^lo^{L-l}}$$ 
Furthermore, $H_S$ has a subspace $\mb{S}_1$ that contains $\ket{\eta}$, composed of $L + 1$ nondegenerate eigenstates. These states are resolved by at least $\Delta$, with $\Delta^{-1} = O\left( nL^5\right)$. The state $\ket{G} = \ket{0^n0^L} \in \mb{S}_1$ has spectral decomposition
\ba
\ket{G} = \frac{1}{\sqrt{L+1}} \ket{\eta} + \sqrt{\frac{2}{L+1}} \sum_{k = 1}^{L} \cos\left(\frac{k \pi}{2(L+1)} \right)\ket{k}
\label{Gdef}
\ea
where $\ket{k}$ is the $k$th excited state in $\mb{S}_1$. Finally, $\mb{S}_1$ is also invariant under the single qubit operator $T_S = \Omega_0\mb{1}_{n} \otimes \ketbrad{0} \otimes \mb{1}_{L-(n+1)}$, which satisfies $\mb{S}_1 T_S \mb{S}_1 = \Omega_0 \ketbrad{G}$.
\end{theorem}

Using the results of Theorem \ref{thm3}, Theorem \ref{thm5} implies that we may produce the history state of $\mb{U}$ using total simulation time $T =\frac{1}{\omega} \cdot O\left(\frac{n L^{9}\sqrt{L}\log(L)}{\epsilon} \right)$, and a cost of $||H_S ||T =  O\left(\frac{n L^{10}\sqrt{L}\log(L) }{\epsilon} \right)$. If we concatenate $\frac{1-\epsilon}{\epsilon}L$ identity operations to the definition of $\mb{U}$, we see that $\ket{\eta}$ is then $\epsilon$-close (with respect to the trace-norm) to the state $\mb{U}\ket{0^n}$. We conclude that any computational problem on $n$ qubits that may be solved at $\poly(n)$ cost using standard quantum computation may also be solved by QSC at a cost of $\poly(n)$.

The ground state of $H_S$ can also be produced with the alternative scheme. As seen in the proof of Theorem \ref{thm5}, in the language of that scheme we may define $\ket{0} = \ket{\eta}$, $\ket{1} = \ket{k = 1}$ as any other eigenstate in the low energy subspace, and $\ket{F} = \ket{G} = \ket{0^n0^L}$. In this case, the probabilistic scheme produces $\ket{\eta}$ at an average cost of $ ||H_S|| \mean{T} = O\left(\frac{L^7\sqrt{L}}{\epsilon^{3/2}} \right)$. We could produce the state $\mb{U}\ket{0^n0^L}$ by adding $M$ identities to $\mb{U}$, then measuring if the clock states one of $\ket{1^L0^{L+1}}$ through $\ket{1^{2L}}$. Since the scheme is already probabilistic, the cost of producing $\mb{U}\ket{0^n}$ would only change by a constant multiple factor.\\\\
\noindent {\it Proof of Theorem \ref{thm5}:}\\
 For this we use Kitaev's clock Hamiltonian \cite{Kitaev2002}, which acts on a system of $n+L$ qubits. The first $n$ qubits represent the actual computation, and start in the fiducial state $\ket{0^n}$. The $L$ other qubits are the 'clock' qubits (denoted by $^c$) that keep track of the evolution of the system. In the notation of \cite{Aharonov2007}, assuming a characteristic energy scale $\omega$, the Hamiltonian is
\ba H_S = \omega H_{prop} +  \Delta_1 H_{input} + 2 \omega  H_{clock}\ea
where
\ba H_{input} = \sum_{i = 1}^n \ketbrad{1}_i \otimes \ketbrad{0}^c_1\ea
 ensures that the initial clock state $\ket{0^L}^c$ is associated with the initial computational state $\ket{0^n}$,
\ba H_{clock} = \sum_{l =1}^{L-1} \ketbrad{01}^c_{l,l+1}\ea
gives an energy cost for not being a valid clock state $\ket{1^l 0^{L-l}}^c$, and 
\ba
H_{prop} = \frac{1}{2}\sum_{l=1}^LH_l
\ea
where for $1<l<L$
\ba 
\nonumber H_{l} & =&  \mb{1}\otimes 
\left(\ketbrad{100}^c_{l-1,l,l+1} + \ketbrad{110}^c_{l-1,l,l+1}  \right) \\
\nonumber &-& U_l \ketbradt{110}{100}^c_{l-1,l,l+1} - U_l^\dagger \ketbradt{100}{110}^c_{l-1,l,l+1}
\ea
correspond to the tracked evolution of the computational bits. $H_1$ and $H_L$ are similarly defined, but with clock qubits $0$ and $L+1$ omitted. In the above notation, the subscripts refer to action on a specific qubit, and imply that other qubits are left unchanged by the operator.

Notice that $H_{input}$, $H_{prop}$ and $H_{clock}$ are each positive semidefinite.  Define $\mb{S}_{legal}$ as the null space of $H_{clock}$, which is compose of states the form $\ket{\cdot}\ket{l}^c = \ket{\cdot}\ket{1^l0^{L-l}}^c$. Since $H_{clock}$ commutes with $H_{prop}$ and $H_{input}$, we see that eigenstates of $H_S$ in the complement of $\mb{S}_{legal}$ have energy at least $2 \omega$. We now reduce our analysis to $\mb{S}_{legal}$, since our desired subspace $\mb{S}_1$ will be contained in $\mb{S}_{legal}$ and will describe energies less than $2 \omega-\Delta$. As in \cite{Kempe2004}, within $\mb{S}_{legal}$ we apply the change of basis $W = \sum_{l = 0}^L U_l U_{l-1} ... U_1 \otimes \ketbrad{l}^c$. $H_{prop}$ and $H_{input}$ are then mapped to
\ban
\tilde H_{input} & =& W^\dagger \left(\sum_{i = 1}^n \ketbrad{1}_i\otimes \ketbrad{0}^c\right) W\\ 
&  =& \sum_{i = 1}^n \ketbrad{1}_i\otimes \ketbrad{0}^c\\
\tilde H_{prop} &=& \frac{1}{2}\mb{1}\otimes\sum_{l=1}^{L}\Big( \ketbrad{l-1}^c + \ketbrad{l}^c \\
& \quad & \quad \quad \quad - \ketbradt{l-1}{l}^c - \ketbradt{l}{l-1}^c\Big) 
\ean
Since it is tridiagonal, it is not hard to show that $\tilde H_{prop}$ has eigenstates of the form
\ban
\ket{x}\ket{\eta_k}^c = 
   \ket{x}\left(\sqrt{\frac{2-\delta_{0 k}}{L+1}} \sum_{l=0}^L \cos\left(\frac{ (l+1/2) k \pi}{L+1} \right)\ket{l}^c \right) 
\ean
with eigenvalue $E_k = 1 - \cos\left(\frac{k \pi}{L+1} \right)$ for $0\leq k \leq L$. We see that these eigenvalues are separated by at least $E_{10}= E_1 - E_0 = 2 \sin^2(\frac{\pi}{L+1}) $, which is greater than $\frac{ \pi^2}{2(L+1)^2}$ for $L\geq 1$. Furthermore, the largest eigenvalue $E_L$ is bounded by $1 - \cos\left(\frac{L \pi}{L+1} \right)$, so $\mb{S}_{legal}$ is separated in energy from its complement by at least $\omega \cdot \frac{1}{O(L^2)}$, thereby justifying our reduction to $\mb{S}_{legal}$.

Within $\mb{S}_{legal}$, for each logical state $\ket{x}$ on the $n$ computational qubits, define the space $\mb{S}_x^*$ by its projector $\mb{S}_x^* = \ketbrad{x}\otimes \sum_{l=0}^L \ketbrad{l}^c$. $\mb{S}_x^*$ commutes with $\tilde H_{input}$ and $\tilde H_{prop}$, so each forms an invariant subspace of $W^\dagger H_S W$. Furthermore, within $\mb{S}_x^*$ we see that
\ban
\mb{S}_x^* \tilde H_{input} \mb{S}_x^*  = N(x) \ketbrad{x}\otimes \ketbrad{0}^c
\ean
where from the previous equation 
\ba
\label{zerot}
\ket{0}^c = \sum_k \sqrt{\frac{2-\delta_{0 k}}{L+1}} \cos \left(\frac{  k \pi}{2(L+1)} \right) \ket{\eta_k}^c
\ea
and $N(x)$ is the number of $1$'s in the binary expression for $x$. 

Within $\mb{S}_x^*$, the Hamiltonian is then
\ban
\mb{S}_x^* W^\dagger H_S W \mb{S}_x^* &=& \mb{S}_x^* W^\dagger (\omega H_{prop}+\Delta_1 H_{input}) W \mb{S}_x^* \\
&=&  \ketbrad{x}\otimes \left(  \omega \sum_{k} E_k \ketbrad{\eta_k}^c + N(x) \Delta_1 \ketbrad{0}^c\right)
\ean
Scaling $\Delta_1/\omega$ as $h \cdot \frac{ \pi^2}{2 n (L+1)^2} < h\cdot E_{10}/n $, we see that $||\Delta_1 \tilde H_{input}|| < h \cdot \omega E_{10}$, where  $\omega E_{10}$ is the minimum eigenvalue spacing of $\tilde H_{prop}$. Hence within each space $\mb{S}_x^*$ we may treat $\Delta_1 \tilde H_{input}$ as a perturbation to $\omega \tilde H_{prop}$, which for small $h$ is well approximated by first order perturbation theory:
 \ban
E_k &\rightarrow& E_k+  \Delta_1 \bra{ \eta_k} \tilde H_{input} \ket{\eta_k}(1 + O(h)) \\
& = & E_k + \Delta_1\frac{ N(x)(2-\delta_{0 k})}{L+1}\cos^2\left(\frac{  k \pi}{2(L+1)} \right)(1 + O(h))
\ean
Since $N(0) = 0$, we see then that the eigenstates of $ \tilde H_{input} + \tilde H_{prop}$ in $\mb{S}_{0^n}^*$ are separated in energy from the other invariant subspaces by at least $\Delta_1\frac{1}{L+1}\cos^2\left(\frac{  L \pi}{2(L+1)} \right)(1 + O(h))$. We conclude that the eigenstates in $\mb{S}_{0^n}^*$ are spectrally resolved by the gap $\Delta$, with
\ban
\Delta^{-1} &=& \Delta_1^{-1} (L+1) \cos^{-2}\l(\frac{  L \pi}{2(L+1)} \r)(1 + O(h)) \\
& = &\frac{1}{\omega} \cdot O\left(n L^5 \right)
\ean

With this information, we may define $\mb{S}_1$ by its projector
\ban
\mb{S}_1 = W \mb{S}_{0^n}^* W^\dagger
\ean
$\mb{S}_1$ is spanned by the eigenstates 
\ban
 \ket{k} &=& W \ket{0^n}\ket{\eta_k}\\
& =& \sqrt{\frac{2-\delta_{0 k}}{L+1}} \sum_{l = 0}^L \cos\left(\frac{(l+1/2) \pi k}{L+1} \right)U_l U_{l-1} ... U_1\ket{0^n} \ket{l}^c
\ean
 with energy $\omega_k = \omega\cdot (1 - \cos\left(\frac{k \pi}{L+1} \right))$, where $\ket{\eta}$ is the $k = 0 $ state. Furthermore, these eigenstates are non-degenerate, and gapped by $\Delta$ as defined above. By the definition of the eigenstates and \eqref{zerot}, we conclude that the state $\ket{G} = \ket{0^n}\ket{ 0^L}^c = \ket{0^n}\ket{l=0}^c$ is contained in $\mb{S}_1$, and that it satisfies \eqref{Gdef}. We note that
\ban
\ketbrad{0}_1^c \cdot \ket{k} &=& \sqrt{\frac{2-\delta_{0 k}}{L+1}} \cos\l(\frac{ \pi k}{2(L+1)}\r) \ket{G}
\ean
so the operator $T_S = \ketbrad{0}^c_1$ leaves the space $\mb{S}_1$ invariant, with $\mb{S}_1 T_S \mb{S}_1 = \ketbrad{G}$. 
\begin{flushright}
$\square$
\end{flushright}


\end{document}